\documentclass[useAMS,usenatbib]{mn2e}
\usepackage{subfigure}
\usepackage{amsmath}
\usepackage{amssymb}
\usepackage{bm}
\usepackage{dcolumn}
\usepackage{graphicx}
\usepackage{color}
\usepackage{ulem}

\newcommand       \apj          {ApJ}
\newcommand       \apjl         {ApJL}
\newcommand       \aap          {A\&A}
\newcommand       \nat          {Nature}
\newcommand       \mnras        {MNRAS}

\newcommand       \prd      {Phys.~Rev.~D.~}

\newcommand      \physrep {Phys.~Rep.}

\makeatletter
\def\tbcaption{\def\@captype{table}\caption}
\def\figcaption{\def\@captype{figure}\caption}
\makeatother

\title[Mass ejection and radio signals from NS mergers]
  {Mass ejection from neutron star mergers: different components and expected radio signals}
\author[K.~Hotokezaka and T.~Piran]
  {Kenta Hotokezaka\thanks{E-mail: kenta.hotokezaka@mail.huji.ac.il} and Tsvi Piran
\\
  Racah Institute of Physics,~The Hebrew~University of Jerusalem,~Jerusalem,~91904,~Israel\\
  }
\date{8 January 2015}

\pagerange{\pageref{firstpage}--\pageref{lastpage}} \pubyear{2015}

\def\LaTeX{L\kern-.36em\raise.3ex\hbox{a}\kern-.15em
    T\kern-.1667em\lower.7ex\hbox{E}\kern-.125emX}

\begin{document}

\label{firstpage}

\maketitle

\begin{abstract}
In addition to producing a strong gravitational signal, a short gamma-ray burst~(GRB), and
a compact remnant, neutron star mergers eject significant masses at significant kinetic energies. 
This mass ejection takes place via dynamical mass ejection and a GRB jet
but other processes have also been suggested: a shock-breakout material,  a cocoon resulting from the interaction of the jet with other ejecta, and 
viscous and neutrino driven winds from the central remnant or the accretion disk.  
The different components of the ejected masses include up to a few percent of a solar mass, some of which is ejected at relativistic velocities.
The interaction of these ejecta with the surrounding interstellar medium will produce a long lasting radio flare, in a similar way to GRB afterglows or to radio supernovae. 
The relative strength of the different signals depends strongly on the viewing angle. 
An observer along the jet axis or close to it will detect a strong signal at a 
few dozen days from the radio afterglow (or the orphan radio afterglow)
produced by the highly relativistic GRB jet. 
For a generic observer at larger  viewing angles,  the dynamical ejecta, whose contribution
peaks a year or so after the event, will generally dominate. Depending on  the observed frequency and  the external density, 
other components may also give rise to a significant contribution. 
We also compare these estimates with the radio signature of the short GRB 130603B.
The radio flare from the dynamical ejecta might be detectable with the EVLA and the LOFAR 
for the higher range of external densities $n\gtrsim 0.5{\rm cm^{-3}}$. 
\end{abstract}


\section{Introduction}
\label{sec:Introduction}

A binary neutron star~(ns$^2$)
merger is one of the most promising targets
of ground-based gravitational-wave~(GW) interferometers, such as Advanced LIGO,
Advanced Virgo, and KAGRA~\citep{ligo2014,virgo2011CQG_v2,kagra2013PRD}. 
The expected event rate of
ns$^2$ mergers is $0.4$~--~$400 {\rm~yr^{-1}}$~\citep{abadie2010CQG}.
Most of these events will be just above or just 
below the detection threshold. 
Observations of an electromagnetic counterpart  will confirm 
the validity of these GW signals, increasing significantly the potential detection rate 
and detection confidence~\citep{kochanek1993ApJ}. In addition,
an electromagnetic counterpart will enable the localization of the sources and  
the identification of their host galaxies and  their redshifts, enhancing significantly the potential
information from this event. 
Finally any electromagnetic counterpart 
will provide invaluable information on the physics of the merger process.

\begin{table*}
\begin{minipage}{18cm}
\begin{center}
\caption[]{
  Summary of mass ejection in different forms.\\
}
 \begin{tabular}{lcccccccc} \hline \hline
  & Mass~[$M_{\odot}$] & Kinetic Energy $E$~[erg] &Fiducial $E$~[erg]& Average~$\beta \Gamma$& Fiducial $\beta \Gamma$& Average~$Y_{e}$ & Reference \\ \hline
Dynamical ejecta\footnote{This component is composed by the tidal tail and shocked component. 
The main difference between them is the value of $Y_{e}$. The tidal component can have a lower $Y_{e}$.}
 & $10^{-4}$~--~$10^{-2}$ & $10^{49}$~--~$10^{51}$& $5\cdot 10^{50}$& $0.1$~--~$0.3$ & 0.2& $0.01$~--~$0.4$& [1]\\
 GRB jet & $\lesssim 10^{-8}$ & $10^{47}$~--~$10^{50.5}$ & $10^{48}$  & $> 30$& --&-- & [2]\\ 
Cocoon & $10^{-6}$~--~$10^{-4}$ & $10^{47}$~--~$10^{50.5}$ & $10^{48}$ & $0.2$~--~$10$ & $0.3$ &-- & [3]\\ 
Shock breakout  &$10^{-6}$~--~$10^{-4}$ & $10^{47}$~--~$10^{49.5}$ & $10^{48.5}$ & $1$ & $1$ &--  & [4]\\
Wind\footnote{This component includes two cases depending on whether  the remnant  is a black hole or a neutron star. 
In the former the wind is just from the surrounding disk while in the latter it arises from the neutron star as well.
The wind from the remnant neutron star has a higher value of $\beta \Gamma$ and $Y_{e}$. 
Note that the fiducial value used is an optimistic one.}
 & $10^{-4}$~--~$5\cdot 10^{-2}$ &$10^{47}$~--~$10^{50}$ & $10^{50}$ &$0.03$~--~$0.1$  & $ 0.07$& $0.2$~--~$0.4$ & [5]\\
\hline \hline\\ 
\label{tab1}
\end{tabular}
\end{center}
\vspace{-0.8cm}
{\scriptsize References;\\
$[1]$ \cite{goriely2011ApJ,korobkin2012MNRAS,hotokezaka2013PRDa,bauswein2013ApJa,rosswog2013RSPTA,piran2013MNRAS,wanajo2014ApJ},\\
$[2]$ \cite{nakar2007,wanderman2014},\\
$[3]$ \cite{nagakura2014ApJ,murguia2014ApJ},\\
$[4]$ \cite{kyutoku2014MNRAS,metzger2015MNRAS},\\
$[5]$ \cite{dessart2009ApJ,fernandez2013MNRAS,metzger2014MNRAS,perego2014MNRAS,just2014,fernandez2015MNRAS}.
}
\vspace{-0.5cm}
\end{minipage}
\end{table*}

The detection horizon distance  will extend up to a few hundred Mpc. The size of the GW-sky localization error box will depend on the number 
of detectors used and between a few tens and thousands Milky-Way size galaxies will reside within this error box~(see e.g.,
\citealt{nissanke2011ApJ}; \citealt{fairhurst2011CQG}). Follow-up observations will be a challenging task, even for a search limited to these galaxies.
Clearly, a good understanding of the expected electromagnetic signals is essential to detect an  electromagnetic counterpart
\citep{metzger2012ApJ,nissanke2013ApJ,kanner2013ApJ,kasliwal2014ApJ,bartos2014MNRAS}.

Ns$^2$ mergers have been recognized as the possible progenitors of short
gamma-ray bursts~(GRBs)~(\citealt{eichler1989Nature}; \citealt{nakar2007}) and 
short GRBs and their afterglows are one of the most attractive electromagnetic 
counterparts of GW events. However, 
GRBs and their early afterglows are believed to be highly beamed with
a half-opening angle $\theta_{j}\approx 10^{\circ}$~\citep{fong2014ApJ}. 
This results in about $5\%$~probability\footnote{Note that these estimates take into account  that the GW horizon is larger in the direction of the GRB
hence the chance of coincidence with a short GRB is larger than the beaming fraction.} to  detect a short GRB in coincidence with the GW
signal~(\citealt{shutz2011CQG}; \citealt{nissanke2011ApJ}; \citealt{seto2015MNRAS}).
Even if the viewing angle is larger than the jet  opening angle,
off-axis afterglows, called orphan afterglows, can be observed at late times
when the  relativistic jet slows down and its emission is less beamed. 
As the frequency of the peak flux of afterglows decreases with time, at sufficiently small viewing angles
the off-axis afterglows in the optical to radio bands can be a good potential candidate of 
electromagnetic counterparts to GW events~(\citealt{vaneerten2011ApJ}; \citealt{metzger2012ApJ}). 

In addition to the GRB beamed emission and its late more isotropic orphan afterglow,
electromagnetic waves will be emitted  quasi isotropically at different stages
from material that is ejected during the merger.
Most notable one is a macronova (also called kilonova), an  optical--infrared transient
 driven by the radioactive decay of the heavy nuclei synthesized
in the ejecta~(\citealt{li1998ApJ}; \citealt{metzger2010MNRAS};
\citealt{kasen2013ApJ}; \citealt{barnes2013ApJ},;
\citealt{tanaka2013ApJ}; \citealt{grossman2014MNRAS}). 
Recently, the {\it Hubble Space Telescope} detected a near infrared bump
at 9~days after the {\it Swift} short GRB 130603B~(\citealt{tanvir2013Nature};
\citealt{berger2013ApJ}), which is consistent with the theoretical expectation of
macronovae.
While the identification hinges on a single data point
if correct this is the first observational evidence for a significant 
mass ejection with a high velocity from a ns$^2$ merger.

Synchrotron radiation of  electrons accelerated in
shocks formed between the (mildly)~relativistic ejecta and the interstellar
medium~(ISM) is a second  electromagnetic counterpart~\citep{nakar2011Nature}. 
This  emission can last up to a few years and peaks in the radio band. 
All the ejected material will contribute to this emission, but 
different components with different velocities will contribute at different timescales, at
different frequencies, and at different intensities. 
The rise time and the peak  flux depend on the density of the
ISM surrounding the merger but for modest densities the radio
signals can be observed up to the detector horizon~\citep{piran2013MNRAS}.

Recent studies have shown that mass ejection from mergers is driven by several different processes. 
The most robust one, that appears in numerical merger simulations, is 
the  dynamical mass ejection.  If mergers are accompanied by GRBs then clearly relativistic GRB jets are another component.
Other mass ejection mechanisms that have been proposed are merger shock-breakout material,
viscous/neutrino/magnetically driven winds, and a possible cocoon that forms when the jet propagates within the other  components of the ejecta.

The  different components have different masses and 
kinetic energies. Their characteristics 
depend on the nature of the progenitors, in particular on their relative sizes, 
on the nature of the merger remnant, which could be either 
a black hole or a massive neutron star~(MNS),
as well on the, unknown yet, neutron star matter equation of state
(see, e.g., \citealt{hotokezaka2013PRDa,bauswein2013ApJa}).
The different components will interact with each other and these interactions will affect their dynamics
~\citep{bucciantini2012MNRAS,nagakura2014ApJ,murguia2014ApJ},
possibly producing electromagnetic signatures
(e.g.,~\citealt{bucciantini2012MNRAS,zhang2013ApJ,metzger2014MNRASmag,nakamura2014ApJ,
rezzolla2014,ciolfi2014,kisaka2014}).

In this paper we examine the long lasting radio emission arising from the different components of the ejecta. 
The structure of the paper is as follows: 
We summarize in Sec.~\ref{sec:Components} the properties of
the different components of the ejecta.
In Sec.~\ref{sec:radio}  we calculate the 
expected long-lasting radio flares produced by the interaction
of the different components of the ejecta with the ISM. 
In Sec.~\ref{sec:grb} we compare these estimates with  the radio signature of
the short GRB 130603B. Finally, in  Sec.~\ref{sec:conc}, we summarize our results and their possible
implications on the detection of radio signals accompanying mergers. 

\section{Different component of ejecta and their properties}
\label{sec:Components}

\begin{figure*}
\includegraphics[bb=0 0 720 540,width=85mm]{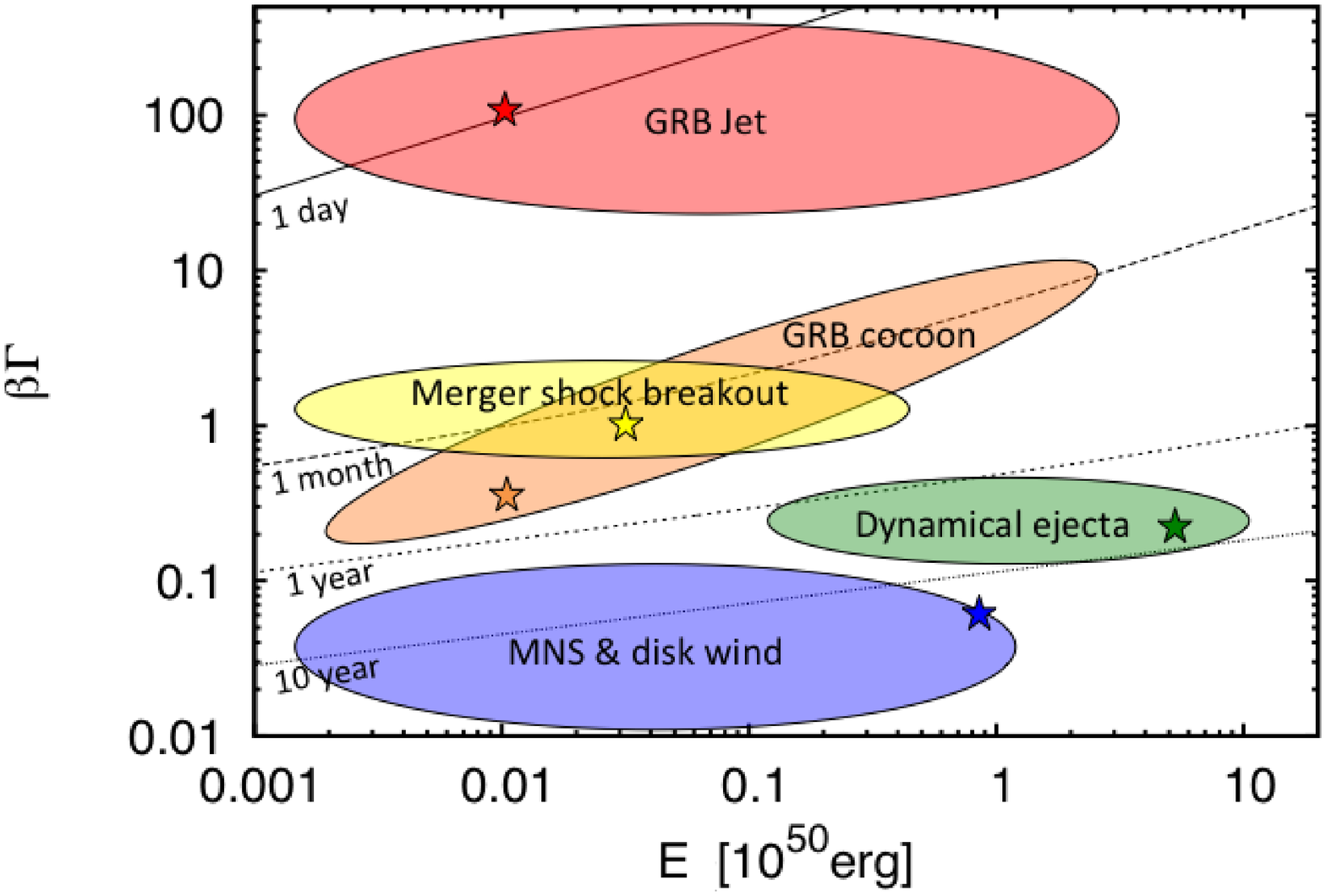}
\includegraphics[bb=0 0 720 540,width=85mm]{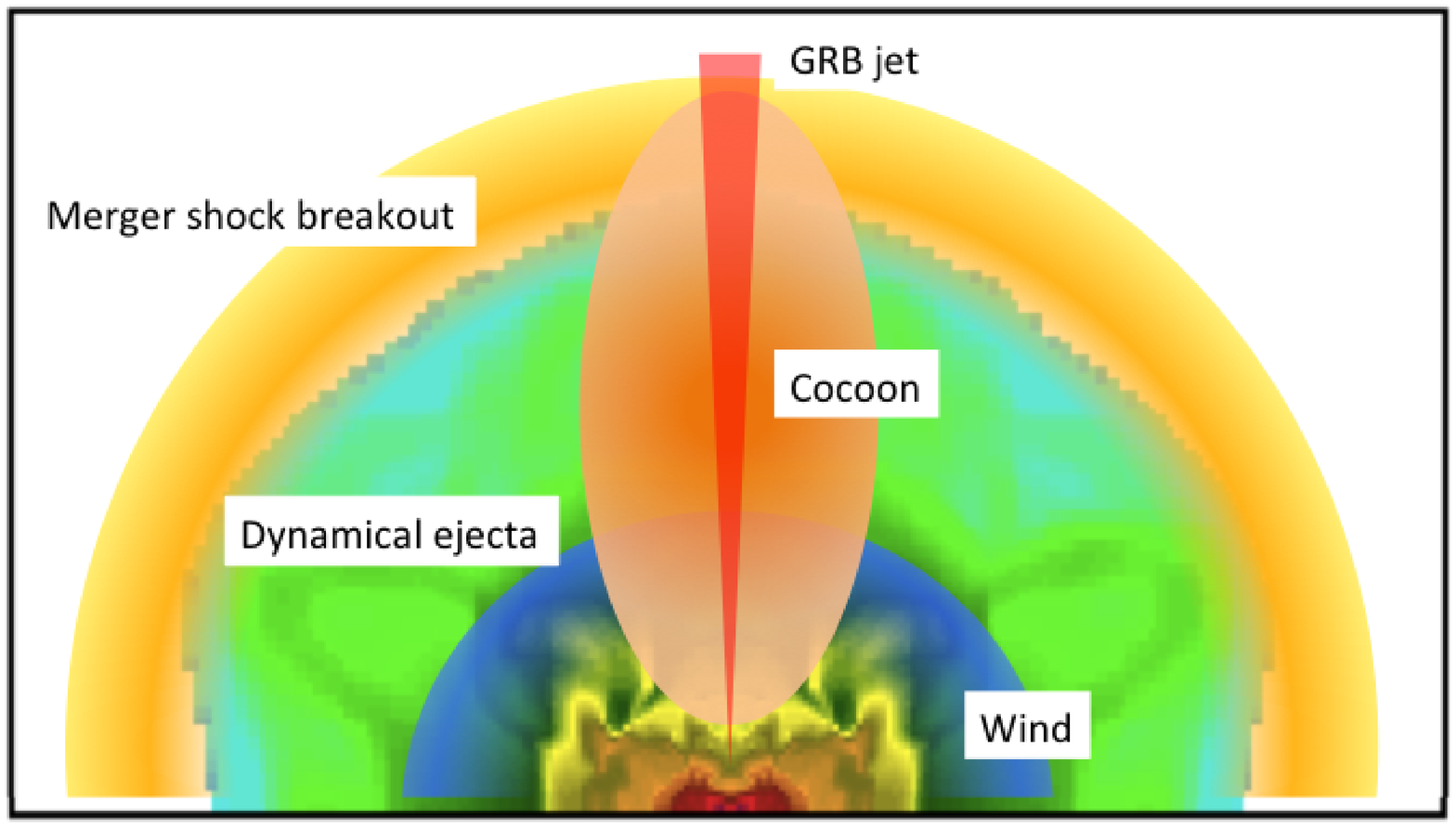}
\caption{Left panel: the kinetic energy and the four-velocity of the different components of the ejecta.
Also marked are 
the deceleration timescales of Eq.~(\ref{tdec1}) assuming an external density of $1~{\rm cm^{-3}}$.
The star in each component shows the fiducial model. 
Right panel: a schematic picture of the morphology of the different components of the ejecta
on the meridional plane. The distribution of the dynamical ejecta is
taken from a merger simulation~\citep{hotokezaka2013PRDa}.
Other components are added schematically.}
\label{fig1}
\end{figure*}

As  material is ejected in different processes the different components will have 
different masses, kinetic energies, velocities, and electron fractions.
The first three quantities determine the radio flare signals while all four are
important for macronova estimate. 
Table~\ref{tab1} summarizes the values of these quantities as taken from
the recent literature.
The properties of the different components of the ejecta are also shown in Fig.~\ref{fig1}.
The left panel of the figure depicts the possible range of the kinetic energy, $E$,
and the four velocity,  $\Gamma \beta$. Here $\Gamma$ is a Lorentz factor and $\beta$
is a velocity in units of the speed of light $c$.
Also shown in the figure are the deceleration timescales due to the interaction with the
ISM,
which are discussed later.
This timescale gives the characteristic peak time of the radio flares from
each component. The right panel of the figure shows schematically the expected
morphology of the ejecta.  

In the following, we briefly describe
the properties of the different components. 
In each case we focus on the total mass, energy, and the corresponding velocities.
We also mention the expected distribution of energy as a function of velocity, which is 
essential in order to estimate the radio flares from these components. 
For completeness we also mention the electron fraction $Y_e$. 
This is not needed for the radio estimate but it is a critical quantity that determines 
the composition of the ejected material as well as the heating rate that is essential for macronova estimates.

\subsection{The dynamical ejecta}
Gravitational and hydrodynamical interactions produce the dynamical ejecta. In many senses it is the easiest to calculate 
and as such it is the most robust element. It 
was investigated using Newtonian simulations~(e.g.,
\citealt{davies1994ApJ,ruffert1997A&A,rosswog1999A&A,rosswog2013RSPTA}) and using general
relativistic simulations~(e.g., \citealt{oechslin2007A&A,hotokezaka2013PRDa,bauswein2013ApJa}).
According to these numerical simulations, the mass and kinetic energy of the 
dynamical ejecta are expected to be in the range $10^{-4}\lesssim M_{\rm ej} \lesssim 10^{-2}M_{\odot}$
and $10^{49}\lesssim E \lesssim 10^{51}$~erg, respectively.
The median value of $E$ in the general relativistic simulations
is a few times $10^{50}$~erg.
The properties of the dynamical ejecta are as follows.
 
{\it The tidal ejecta.}
A fraction of the material obtains sufficient angular momentum and is ejected
via tidal interaction due to non-axisymmetry of the gravitational forces.
This matter  is ejected even before the  two stars collide with each other
and it lasts as long as the gravitational field is not axisymmetric
~(about $10$~ms after the merger in the case that the remnant is
a MNS).
This tidal component is mostly ejected into the equatorial plane of the binary within an angle
about $20^{\circ}$~(see e.g., Fig.~17 in \citealt{hotokezaka2013PRDa}).

The electron fraction of the dynamical ejecta and the resulting nucleosynthesis
have been studied in the literature~(e.g.~\citealt{goriely2011ApJ,korobkin2012MNRAS,wanajo2014ApJ}).
The tidally ejected material has initially a low electron fraction $Y_{e}\ll 0.1$  as
this matter does not suffer from shock heating and neutrino irradiation~\citep{wanajo2014ApJ}. 
This is particularly important concerning the possibility that this is the source of 
heavy (high atomic number) $r$-process nuclides, but it is not so relevant for our discussion that is concerned mostly with  the radio flare.
This fraction can increase by  electron neutrino absorption or by  positron absorption.
The tidal component  ejected at late times has higher $Y_{e}$ values. 

{\it The shocked component.}
A shock is formed at the interface of the merging neutron stars.
The shock sweeps up the material in the envelope of the merging neutron stars. 
Furthermore, a shock is continuously produced around the envelope of a remnant MNS
as long as the MNS has radial oscillation. 
As a result, a fraction of the shocked material obtains sufficient energy and 
is ejected from the system. Recent general relativistic simulations
show that this component can dominate over the tidal component
in the case of a nearly equal mass binary~(e.g., \citealt{hotokezaka2013PRDa,bauswein2013ApJa}).
The shocked component is ejected even in the direction of the rotation axis of the binary. 
The average electron fraction of the shocked components is relatively large compared with that 
of the tidal ejecta~\citep{wanajo2014ApJ}.  It may be as large as  
$Y_{e}\sim 0.2$~--~$0.4$  and it will result in a different nucleosynthesis signature.

We take the velocity distribution of the dynamical ejecta
from the result of a numerical relativity simulation of~\cite{hotokezaka2013PRDa}
for a $1.4$--$1.4M_{\odot}$ ns$^2$ merger for the case of 
APR4 equation of state. 
The energy distribution of this model can be approximately described
as $E(\geq \beta)\propto \beta^{-0.5}$ with a cut off at $\beta \simeq 0.4$ and
an average velocity is $\beta\simeq 0.2$,
where $E(\geq \beta)$ is the kinetic energy with a velocity larger than $\beta$.
Note that it is not clear whether the cut off at $\beta \simeq 0.4$
is physical or that it arises just because it is difficult to resolve such a 
small amount of fast material in the numerical simulations. 
For our fiducial model, we use a total kinetic energy of 
$5 \times 10^{50}~{\rm erg}$.

{\it The relativistic shock-breakout component.}
When the shock breaks out from the neutron star surface to the ISM,
it is  accelerated and a fraction of the shocked component can have 
a relativistic velocity with $\beta \Gamma \gtrsim 1$.
\citet{kyutoku2014MNRAS} showed analytically that the kinetic energy of the
relativistic ejecta can be $\sim 10^{47}$~--~$10^{49.5}~{\rm erg}$.
More recently, \citet{metzger2015MNRAS} found that
there is a mildly relativistic component with $\beta\gtrsim 0.8$ 
in a merger simulation of \cite{bauswein2013ApJa}.
This fast component is likely resulted from the acceleration of
a shock emerging from the neutron star surface.   
They found that the mass and  kinetic energy of the fast component with 
$\beta \gtrsim 0.8$ are
$\sim 10^{-5}M_{\odot}$ and $\sim 5\times 10^{48}$erg, respectively.
Because of this large velocity, the radio signature of this component would be different 
from the slower material. 
We denote this component as a ``shock-breakout material'' and
we consider it  separately from the sub-relativistic dynamical ejecta.

Here, we assume that the kinetic energy distribution of this component is 
a simple power-law, 
$E(\geq \beta \Gamma)=10^{48.5}(\beta \Gamma)^{-\alpha}~{\rm erg}$
as the fiducial model.
The value of $\alpha$ varies from $1.1$ for $\beta \Gamma \gg 1$ to
$5.2$ for $\beta \Gamma \ll 1$~\citep{kyutoku2014MNRAS,tan2001ApJ}.  
We set $\alpha$ to be $3$, which is valid around $\beta \Gamma\sim 1$ and
we take into account only the fast component with $\beta \Gamma\geq 1$.

\subsection{The ultra-relativistic jet}

If ns$^2$ mergers are progenitors of short GRBs, they involve relativistic jets.
Assuming the kinetic energy of the jet as
the gamma-ray energy of the prompt emission, the energy of
the jet can be estimated from the observed GRBs.
The minimal and maximal values of the observed isotropic-equivalent
gamma-ray energy for non-Collapsar short GRBs
are $2\times 10^{49}$~erg and $4\times 10^{52}$~erg,
respectively~\citep{wanderman2014}. 
Taking into account the average value of
the measured jet-half opening angles of 
$\theta_{j} \approx 10^{\circ}$~\citep{fong2014ApJ},
the kinetic energy of a relativistic jet is
in the range of $10^{47}$~--~$3\times 10^{50}~{\rm erg}$.
The luminosity function is rather steep and 
there are more  weak GRBs than strong ones.
Hence we consider here a fiducial  GRB jet with a  kinetic energy of $10^{48}~{\rm erg}$
and a jet-half opening angle of $10^{\circ}$.

\subsection{The wind from the merger remnant} 

Some of the debris of the neutron stars form an accretion disk that surrounds
the central remnant.  The mass of this accretion disk is estimated to be in the range of
$10^{-3} \lesssim M_{\rm disk} \lesssim 0.3~M_{\odot}$~(see e.g., \citealt{shibata2006PRD,rezzolla2010CQG,
hotokezaka2013PRDc}).
This accretion disk  produces an outflow
driven by viscous and neutrino heating.
The properties of this outflow depend on the central object as follows.

{\it A black hole with an accretion disk.}
The wind from an accretion disk surrounding a black hole
has been explored, in the context of mergers,  by~\cite{fernandez2013MNRAS,
just2014,fernandez2015MNRAS}. The disk is extremely dense and the accretion rate is huge. 
Initially it is opaque even for neutrinos.  
After $\sim 0.1$~--~$1$~s from the  onset of the merger,
the density and temperature of the accretion disk decrease and
neutrino-cooling becomes inefficient. As a result, a fraction
of the material is ejected isotropically due to the viscous heating
in the accretion disk.
The amount of ejected material is about $5$~--~$20\%$ of the initial disk mass
depending on the $\alpha$-viscosity parameter and on the spin parameter of the
black hole. Increasing these parameters, the fraction of the
ejected mass to the initial disk mass increases.
The average velocity of the ejecta is $0.03 \lesssim \beta \lesssim 0.05$ and 
the expected kinetic energy of the outflow is in the range of
$10^{47} \lesssim E \lesssim 10^{50}$~erg. 
The average electron fraction is $Y_{e}\sim 0.2$~--~$0.3$.

{\it A neutron star with an accretion disk.}
The wind from a  neutron star with an accretion disk
can be divided into three parts; a neutrino-driven wind
from the remnant neutron star itself, a neutrino driven wind
from the accretion disk, and a viscous driven wind from
the accretion disk~\citep{dessart2009ApJ,
metzger2014MNRAS,perego2014MNRAS}.
The neutrino-driven wind expands into relatively high latitudes and it 
has a larger velocity $\beta \lesssim 0.1$ and a higher electron fraction
$Y_{e}\sim 0.4$ than those of the viscous-driven wind.
The amount of material ejected by the neutrino-driven wind 
depends on the lifetime of the central neutron star.
\cite{perego2014MNRAS} showed that the ejected mass
is more than $3\times 10^{-3}M_{\odot}$ with a velocity $\beta\sim 0.06$~--~$0.9$
at 100~ms after the merger in the case of an initial disk mass of $0.17M_{\odot}$.
\cite{metzger2014MNRAS} also showed that 
the amount of the ejected mass and average velocity are
$M_{\rm ej}\sim 10^{-3}M_{\odot}$ and $\beta \sim 0.05$ at 
$100$~ms with an initial disk mass of $0.03M_{\odot}$.
When a MNS does not collapse into a black hole,
about $20\%$ of the initial disk mass may be ejected as neutrino
driven or viscous driven winds~\citep{metzger2014MNRAS}.

In the following, we take the kinetic energy of $10^{50}$~erg with
a single velocity $0.07c$ for the wind from the merger remnant
as the fiducial model.
Note that these values correspond to the most optimistic case.
As shown later even in this case, the expected radio signals are very weak.

\subsection{The cocoon}
The interaction of a GRB
jet with the pre-ejected material such as the dynamical ejecta or 
the wind along the rotation axis would produce a hot cocoon surrounding the jet
~\citep{nagakura2014ApJ,
murguia2014ApJ, rezzolla2014}.
After the jet emerges  from the expanding ejecta,
the cocoon  will break out from the surface of the ejecta and will expand
nearly spherically.
Assuming that the material inside the cone of a jet-half opening angle $\theta_{j}$
is shocked by the jet and forms a cocoon and the deposited energy into the cocoon is $E_{c}$, 
the Lorentz factor of the cocoon can be estimated as
\begin{eqnarray}
\Gamma \approx 1+0.05\left(\frac{M_{\rm ej}(\theta_{j})}{10^{-5}M_{\odot}}\right)^{-1}
\left(\frac{E_{c}}{10^{48}{\rm~erg}}\right),\label{Lc}
\end{eqnarray}
where $M_{\rm ej}(\theta_{j})$ is the ejecta mass within $\theta_{j}$.
As the jet crossing time is comparable with the duration of a short GRB
and the jet energy deposited in the cocoon will be comparable to the jet energy,
we expect that the cocoon energy will be similar to the GRB jet energy. 
For our fiducial value, we take a kinetic energy of $10^{48}$~erg with
a single velocity of $0.3c$ and the cocoon is sub-relativistic.
Note that it becomes relativistic at energies higher than $10^{49}$~erg.

\section{The radio signature}
\label{sec:radio}
The various components of ejecta interact first with each other and ultimately with the ISM. 
This last interaction produces a long-lived blast wave. 
This shock that propagates into the ISM will
enhance magnetic fields and accelerate electrons that will emit synchrotron radiation.
The process is similar to GRB afterglows and to radio emission from some early supernova remnants. 
In this section, we explore the synchrotron radiation from a merger
taking into account the various components of ejecta.
Except for the relativistic  jet we consider  all components of the ejecta 
as spherically symmetric. We discuss the implications of this approximation in Sec.~\ref{sec:results}. 

We assume that the ISM is homogeneous and characterized by an external density $n$. 
The ejecta slows down  with the deceleration timescale given by
\begin{eqnarray}
t_{\rm dec}  =  \left(\frac{3E}{4\pi m_{p}c^{5}n \Gamma_{0} (\Gamma_{0}-1)\beta_{0}^{3}}\right)^{1/3},\label{tdec1}
\end{eqnarray} 
where 
$\Gamma_{0}$ and $\beta_{0}$ are the initial Lorentz factor and the corresponding
initial velocity of the ejecta,
$m_{p}$ is the proton mass.
The values of $t_{\rm dec}$ for the different components of ejecta are shown
in the left panel of Fig.~\ref{fig1}.  
For a mildly or sub-relativistic outflow the deceleration timescale characterizes the observed peak time. 
For an ultra-relativistic beamed jet with a viewing angle $\theta_{\rm obs}>\theta_{j}$, 
we have an orphan afterglow.  Namely, we do not see the highly beamed burst and  early afterglow. But we see the late afterglow when it slows down and its less beamed emission includes our line of sight. 
As this happens when $\Gamma \sim \theta_{\rm obs}^{-1}$,
the peak time in the source frame is around $t_{\rm dec}$ given by 
Eq.~(\ref{tdec1})  with  $\Gamma \sim \theta_{\rm obs}^{-1}$
instead of the initial Lorentz factor.  
Note that for a relativistic outflow the observer time is different from the time in source frame and
it is smaller by a factor of $\Gamma^{-2}$.
Note that for an observed GRB the peak in the radio arises when the observed frequency
equals the  typical synchrotron frequency.

The deceleration timescale~(see the left panel of Fig.~\ref{fig1}) suggests 
three types of the radio flares. 
First, the ultra-relativistic jet produces at early times the radio afterglow, that can be seen only by 
observers along the jet axis or close to it.
Second, the mildly relativistic components, including the cocoon, the  shock-breakout material, and the jet for an observer away from
its axis produce radio flares with a timescale of a few dozen days.  
Finally, the sub-relativistic
dynamical ejecta produces a  late-time radio flare with a timescale
of a few years.

\subsection{Ultra-relativistic beamed jet}
\label{sec:jetafterglow}
The Blandford-Mckee self-similar solution describes the jet dynamics, in the relativistic regime after the energy of the 
ISM swept up by the jet
becomes comparable to the energy of the jet itself.
Once the Lorentz factor of
the jet decreases to $\Gamma \sim \theta_{j}^{-1}$,
the jet expands laterally and approaches a quasi- spherical shape.
To describe the evolution during this sideway expansion phase we adopt a semi-analytic formula 
for a homogeneous jet given by
\cite{granot2012MNRAS}\footnote{We adopt the conical model of \cite{granot2012MNRAS}.
The difference of the afterglow flux between their different models is
a factor $2\sim 3$ during the side-expansion phase.},
which shows a good agreement with the results of a numerical simulation by
\cite{decolle2012ApJ}. 
The observed signal depends strongly on the viewing angle and we consider  
five different viewing angles $\theta_{\rm obs} = (0^{\circ},~30^{\circ},~45^{\circ},~60^{\circ},~90^{\circ})$.

To calculate the synchrotron radio emission we assume, as common 
in GRB afterglows and in radio supernovae modeling~(see e.g., \citealt{sari1998ApJ}), 
that the shock generates magnetic fields 
and accelerates electrons to a power law distribution $N(\gamma)\propto \gamma^{-p}$, where $\gamma$
is the Lorentz factor of an accelerated electron. 
The value of $p$ is estimated as $p \approx 2.1$~--~$2.5$ in
late GRB afterglows and afterglows of low luminosity GRBs
and as $p\approx 2.5$~--~$3$ in typical radio supernovae~\citep{chevalier1998ApJ}.
We assume $p=2.5$. The total energy of the electrons and the magnetic
field intensity are characterized by equipartition parameters:
$\epsilon_{e}$ and $\epsilon_{B}$ that are
the conversion efficiency from the internal energy of the
shock into the energy of the accelerated electrons and magnetic
fields, respectively. 
We set these parameters as $\epsilon_{e}=\epsilon_{B}=0.1$.
These values are consistent with those
evaluated from late radio afterglows in long GRBs~\citep{frail2000ApJ,frail2005ApJ}.
For our purposes
the radio emission is always below the cooling frequency hence the system
has only two characteristic frequencies, the synchrotron frequency of the ``typical" electron
and the self absorption frequency.
We implement the effect of the synchrotron-self absorption
following
 \cite{granot1999ApJb,rybicki1979}.  
Once we determine the local emissivity 
we integrate over the intensity of each line of sight
with an equal arrival time~(see e.g.,~\citealt{sari1998ApJb,granot1999ApJ})\footnote{We do not use the afterglow library of
\cite{vaneerten2011ApJ} which is incorrect below  the
absorption frequency.
Above the absorption frequency our light curves are consistent with
those of \cite{vaneerten2011ApJ}.}.

\subsection{Mildly and sub-relativistic isotropic components.}
\label{sec:mild}

We briefly discuss the simple analytic estimates of the radio signals
for a mildly and sub-relativistic ejecta ~(see \citealt{piran2013MNRAS} for details). 
The hydrodynamics of a mildly and sub-relativistic
blast wave with a kinetic energy $E$ and an initial velocity $\beta_{0}$
expanding into a homogeneous ISM with an external density $n$
can be approximately described by $\beta=\beta_{0}$ until
the deceleration time $t_{\rm dec}$. 
The dynamics approaches
to the Sedov-Taylor self-similar solution after $t_{\rm dec}$.

The synchrotron emission is slow cooling and it is strongly suppressed by the self absorption below the self absorption frequency: 
\begin{eqnarray}
\nu_{a}(t) =
\left\{
\begin{array}{ll}
\nu_{a{\rm ,dec}} \left( \frac{t}{t_{\rm dec}}\right)^{\frac{2}{p+4}}
&~~(t\leq t_{\rm dec}),\\
\nu_{a{\rm ,dec}}\left( \frac{t}{t_{\rm dec}}\right)^{-\frac{3p-2}{p+4}}
&~~(t > t_{\rm dec}),\\
 \end{array}
 \right.
\end{eqnarray}
where
\begin{eqnarray}
\nu_{a{\rm ,dec}} = 1~{\rm GHz}~E_{49}^{\frac{2}{3(p+4)}}
n^{\frac{3p+14}{6(p+4)}}
\epsilon_{B,-1}^{\frac{2+p}{2(p+4)}}
\epsilon_{e,-1}^{\frac{2(p-1)}{p+4}}
\beta_{0}^{\frac{15p-10}{3(p+4)}}.
\end{eqnarray}
These expressions are valid for $\nu_{a}>\nu_{m}$, where $\nu_{m}$
is the synchrotron frequency of electrons with the minimum Lorentz factor.
Here and elsewhere, $Q_{x}$ denotes the value of $Q/10^{x}$ in cgs units.

For $\nu>\nu_{a}$, the peak flux and the peak time can be estimated as
\begin{eqnarray}
F_{{\rm peak},\nu>\nu_{a}(t_{\rm dec})}\approx 0.8~{\rm mJy}~E_{49}n^{\frac{p+1}{4}}\epsilon_{B,-1}^{\frac{p+1}{4}}
\epsilon_{e,-1}^{p-1}\beta_{0}^{\frac{5p-7}{2}}\label{f1} \\ \nonumber
\times \left(\frac{D_{L}}{200~{\rm Mpc}}\right)^{-2}
\left(\frac{\nu}{1.4~{\rm GHz}}\right)^{-\frac{p-1}{2}},
\end{eqnarray}
and
\begin{eqnarray}
t_{\nu > \nu_{a}(t_{\rm dec})}=t_{\rm dec} \approx 40~{\rm day}~E_{49}^{\frac{1}{3}}n^{-\frac{1}{3}}\beta_{0}^{-\frac{5}{3}},\label{t1}
\end{eqnarray}
where $D_{L}$ is the luminosity distance to the source and
we approximate the Lorentz factor as $\Gamma_{0}-1\approx \beta_{0}^{2}$ in Eq.~(\ref{t1}).
The peak flux and its time depend sensitively on the external density,
the kinetic energy, and the initial velocity of the ejecta in the optically thin regime.
  
For $\nu<\nu_{a}$ at $t_{\rm dec}$, the peak flux and peak timescale are
\begin{eqnarray}
F_{{\rm peak},\nu<\nu_{a}}\approx 0.1~{\rm mJy}~E_{49}^{\frac{4}{5}}n^{\frac{1}{5}}\epsilon_{B,-1}^{\frac{1}{5}}
\epsilon_{e,-1}^{\frac{3}{5}}\label{f2} \\
\times \left(\frac{D_{L}}{200~{\rm Mpc}}\right)^{-2}
\left(\frac{\nu}{150~{\rm MHz}}\right)^{\frac{6}{5}},\nonumber
\end{eqnarray}
and
\begin{eqnarray}
t_{\nu < \nu_{a}(t_{\rm dec})}\approx 200~{\rm day}~E_{49}^{\frac{5}{11}}n^{\frac{7}{22}}\epsilon_{B,-1}^{\frac{9}{22}}\label{t2}
\epsilon_{e,-1}^{\frac{6}{11}}\left(\frac{\nu}{150~{\rm MHz}}\right)^{\frac{13}{11}}.\\ \nonumber
\end{eqnarray}
In the optically thick regime, the peak flux and its timescale depend weakly on the external density
and they are independent of the initial velocity of the ejecta. 
The dependence on the energy is also weaker than in the optically thin case.

As the velocity distribution is not uniform, we estimate the emission from each shell of matter
and combined the results. 
For a given distribution of energies as a function of velocity,
we divide the outflow into shells.  An external ISM mass,  $M(R)$, swept up at a radius $R$ can be associated 
with each shell such that this mass slows down the shells:
\begin{eqnarray}
M(R)(c\beta \Gamma)^{2} = E(\geq \beta \Gamma). 
\label{MR}
\end{eqnarray}
Once we solve the implicit Eq.~(\ref{MR}), we determine the observed light curves for 
each shell.
We then combine the contributions of the different shells to obtain the total light curve. 
 In the non-relativistic limit, the ejecta dynamics described by Eq.~(\ref{MR}) is consistent with
the self-similar solution derived by \cite{chevalier1982ApJ} up to a factor of order unity.
In the relativistic limit and the case of $E(\geq \beta \Gamma)={\rm const}$,
it agrees with the Blandford-Mackee self-similar solution again up to a factor of order unity.

\subsection{Numerical result}
\label{sec:results} 

\begin{figure*}
\includegraphics[bb=50 50 410 302,width=85mm]{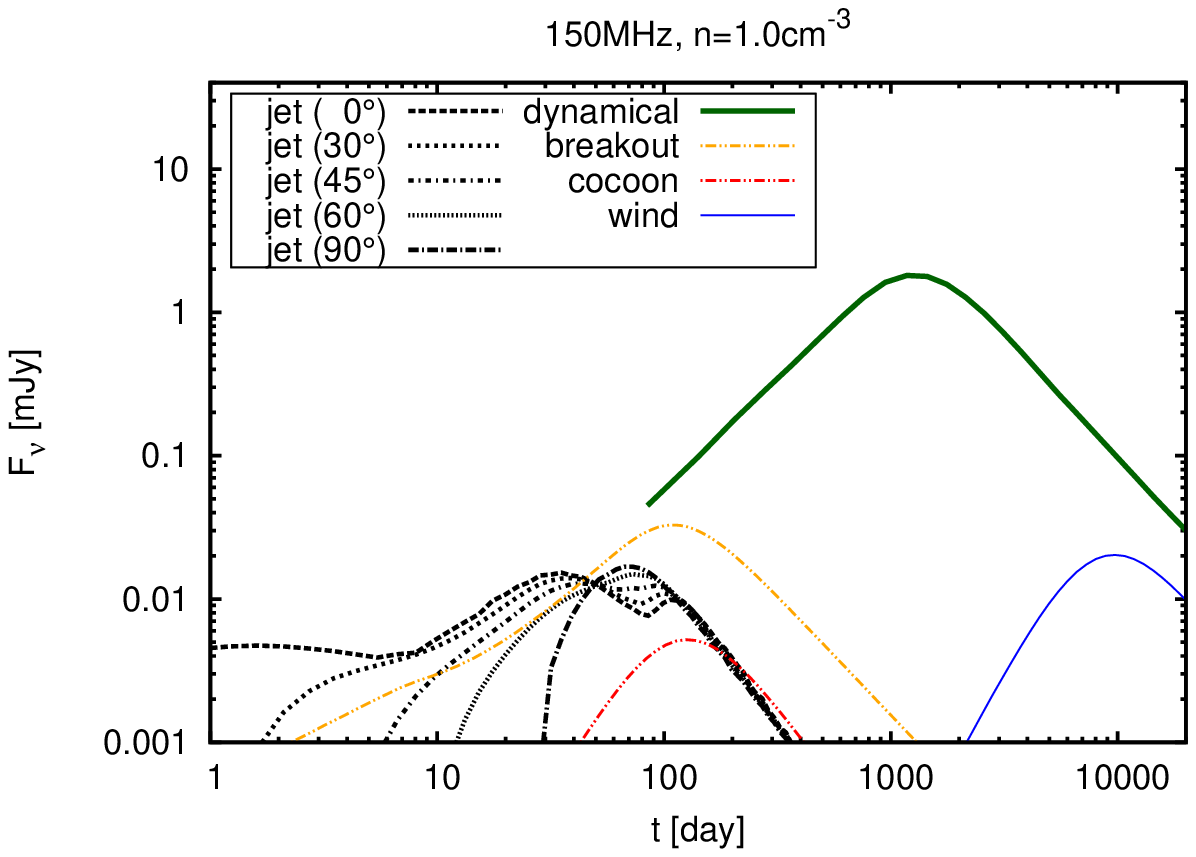}
\includegraphics[bb=50 50 410 302,width=85mm]{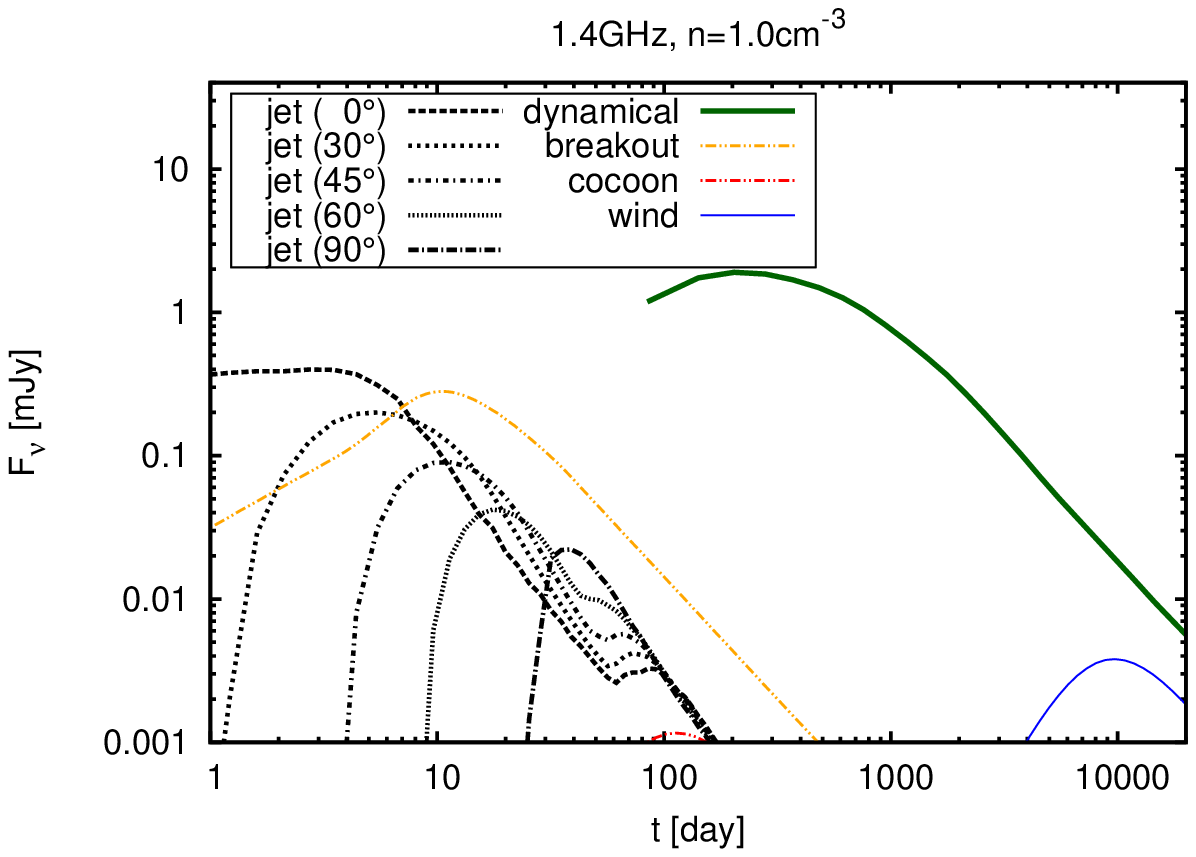}\\
\includegraphics[bb=50 50 410 302,width=85mm]{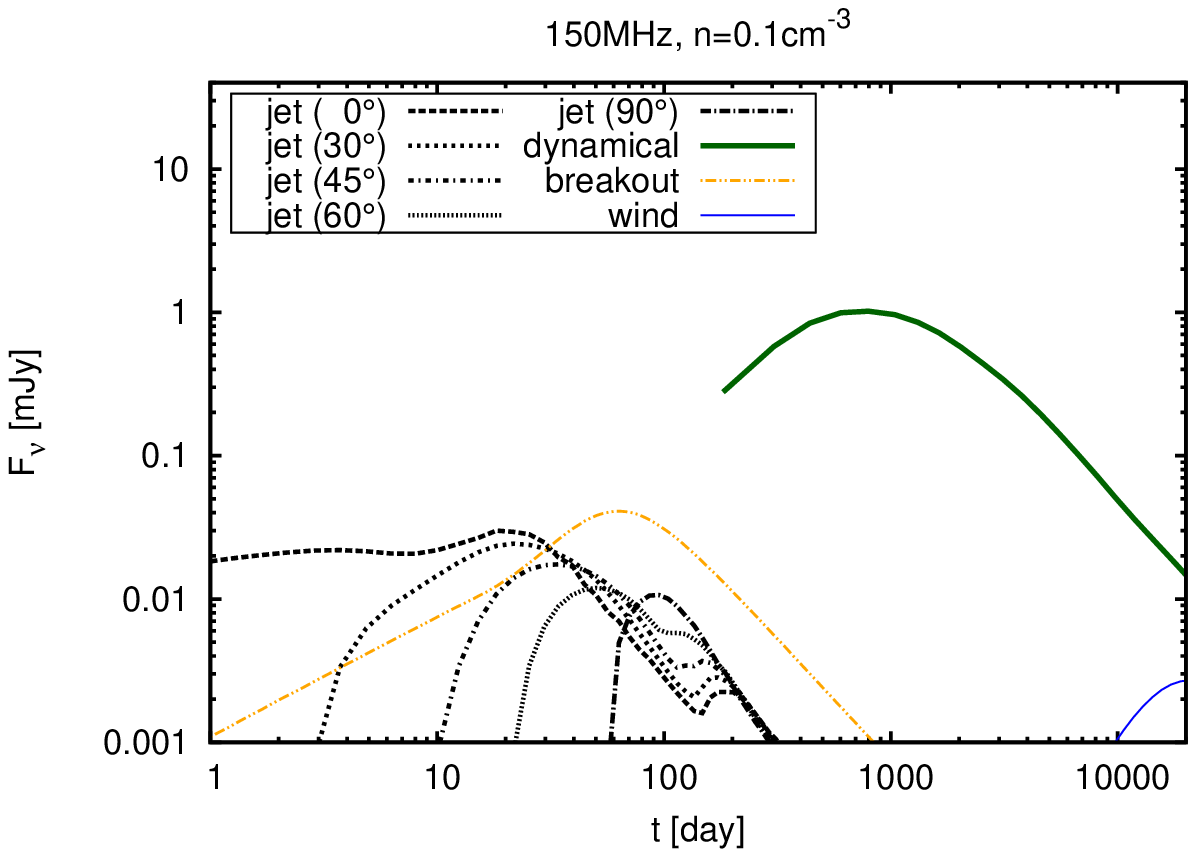}
\includegraphics[bb=50 50 410 302,width=85mm]{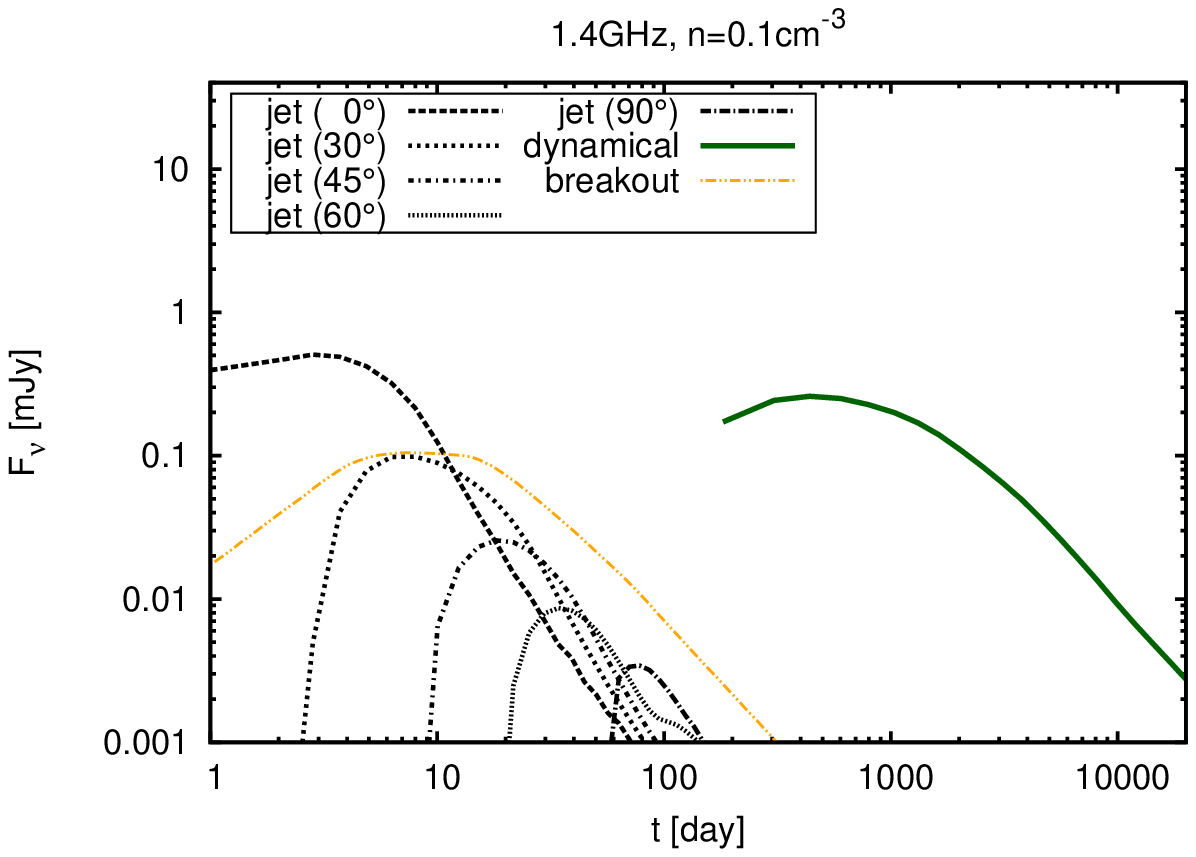}\\
\includegraphics[bb=50 50 410 302,width=85mm]{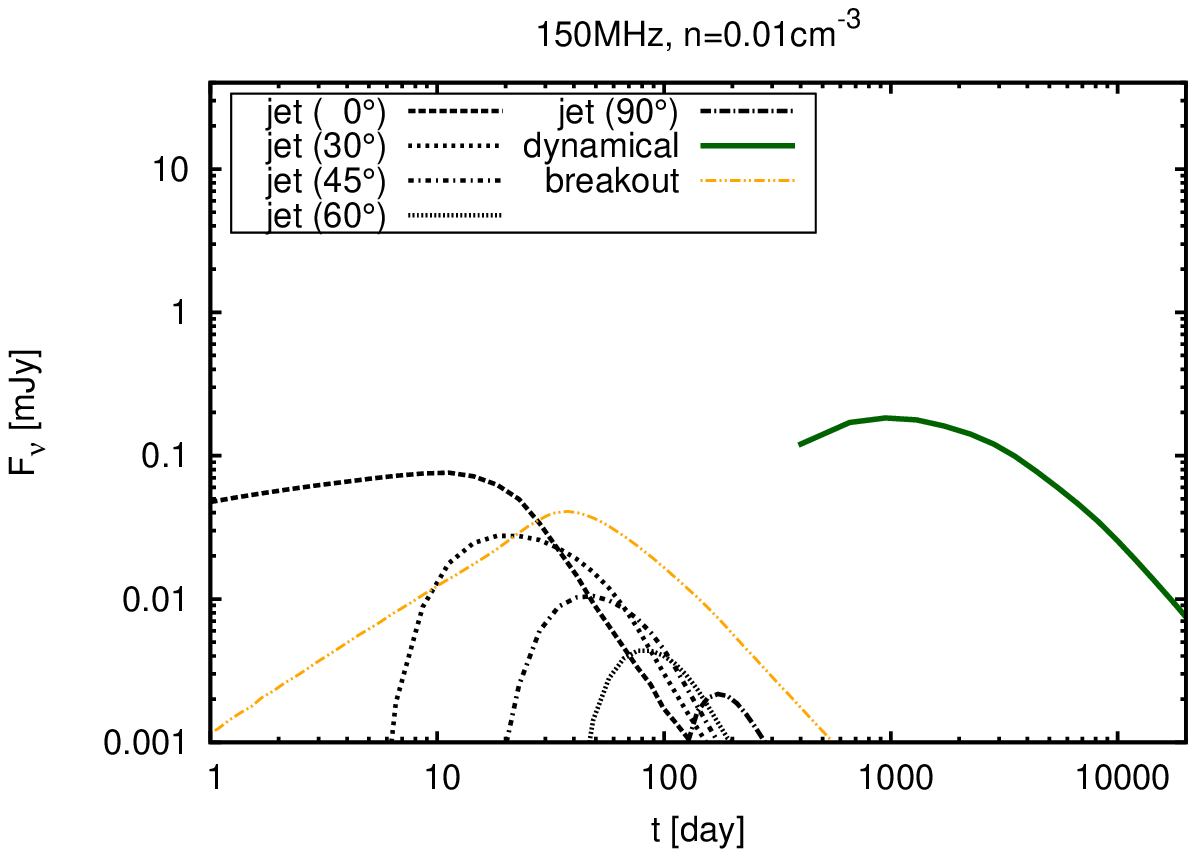}
\includegraphics[bb=50 50 410 302,width=85mm]{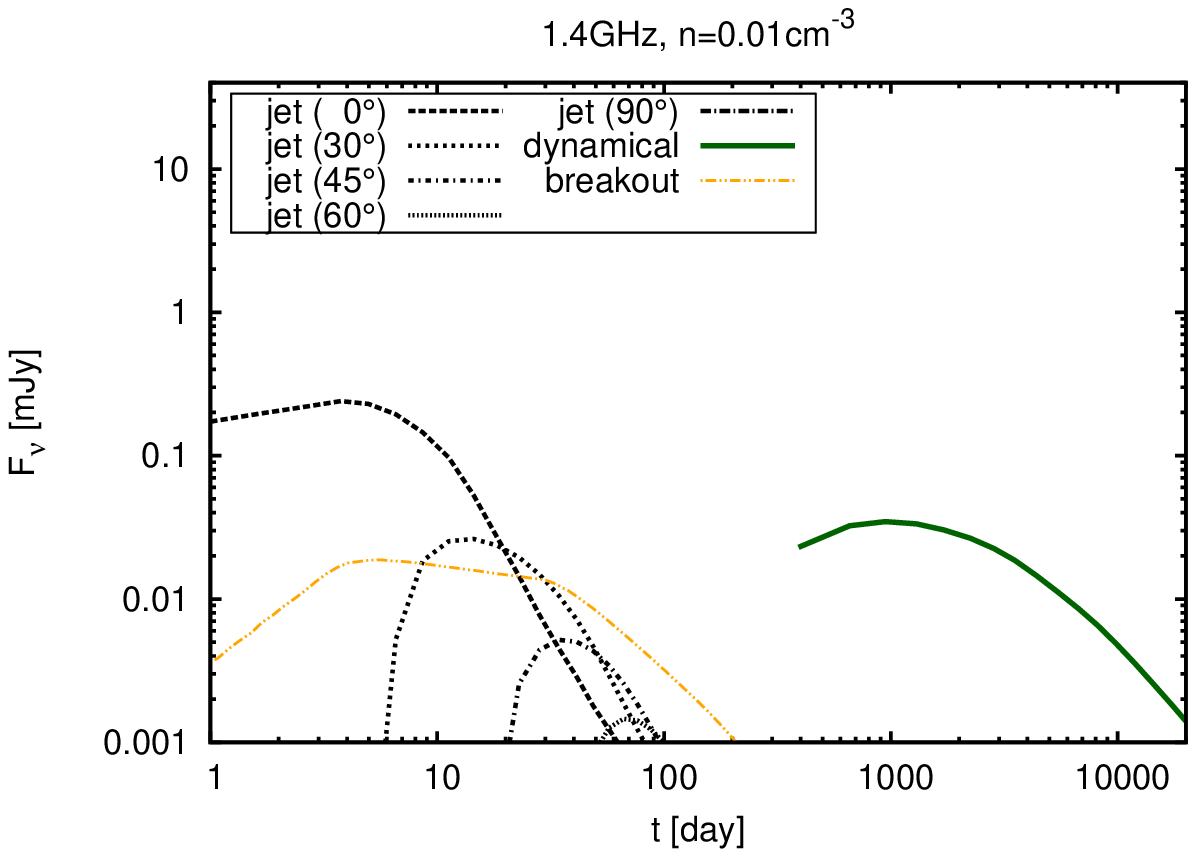}
\caption{
Radio light curves from the different ejecta components for the fiducial model.
Here the energy of the ultra-relativistic jet and the cocoon is set to be $10^{48}$~erg and
the velocity of the cocoon is $0.3c$.
The external density and
the luminosity distance to the source are set to be $0.01$~--~$1~{\rm cm^{-3}}$
and $200$~Mpc. 
The left and right panels show the radio light curves at 150~MHz
and 1.4~GHz, respectively.}
\label{fig2}
\end{figure*}

\begin{figure*}
\includegraphics[bb=50 50 410 302,width=85mm]{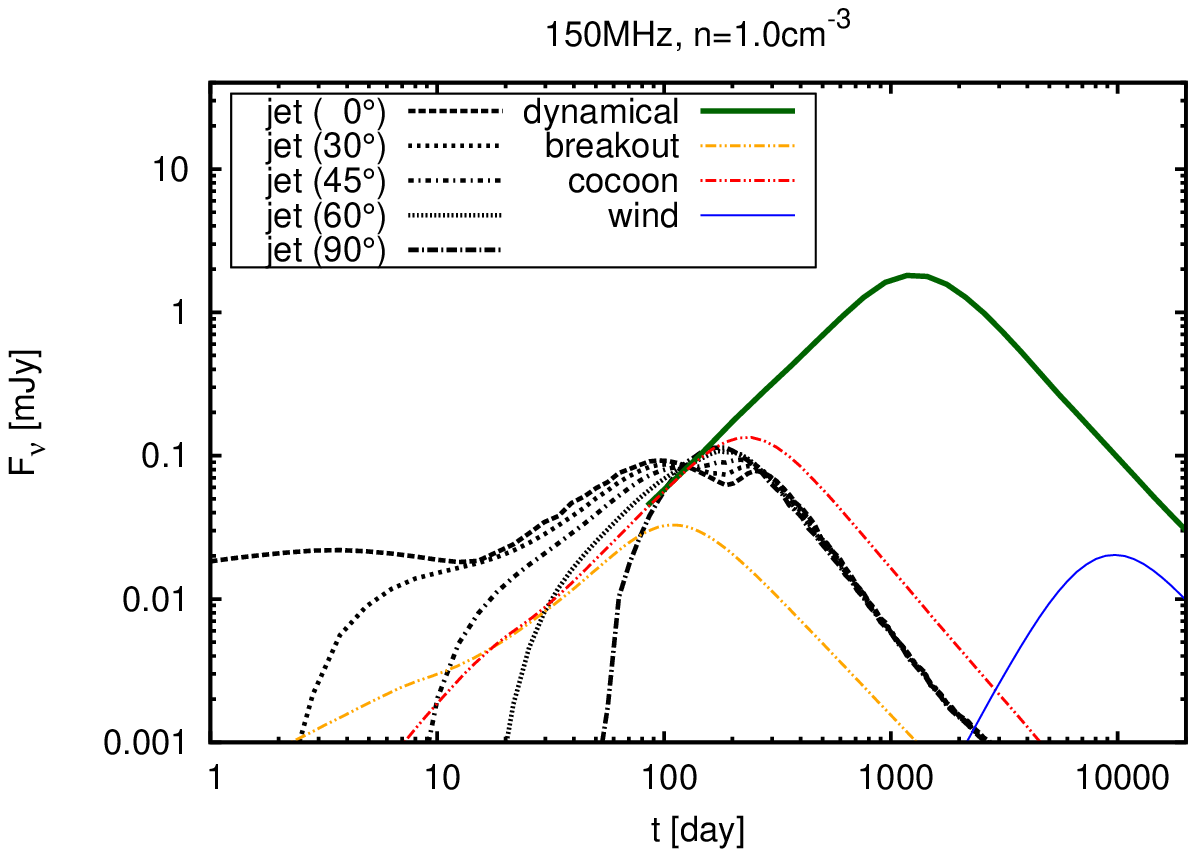}
\includegraphics[bb=50 50 410 302,width=85mm]{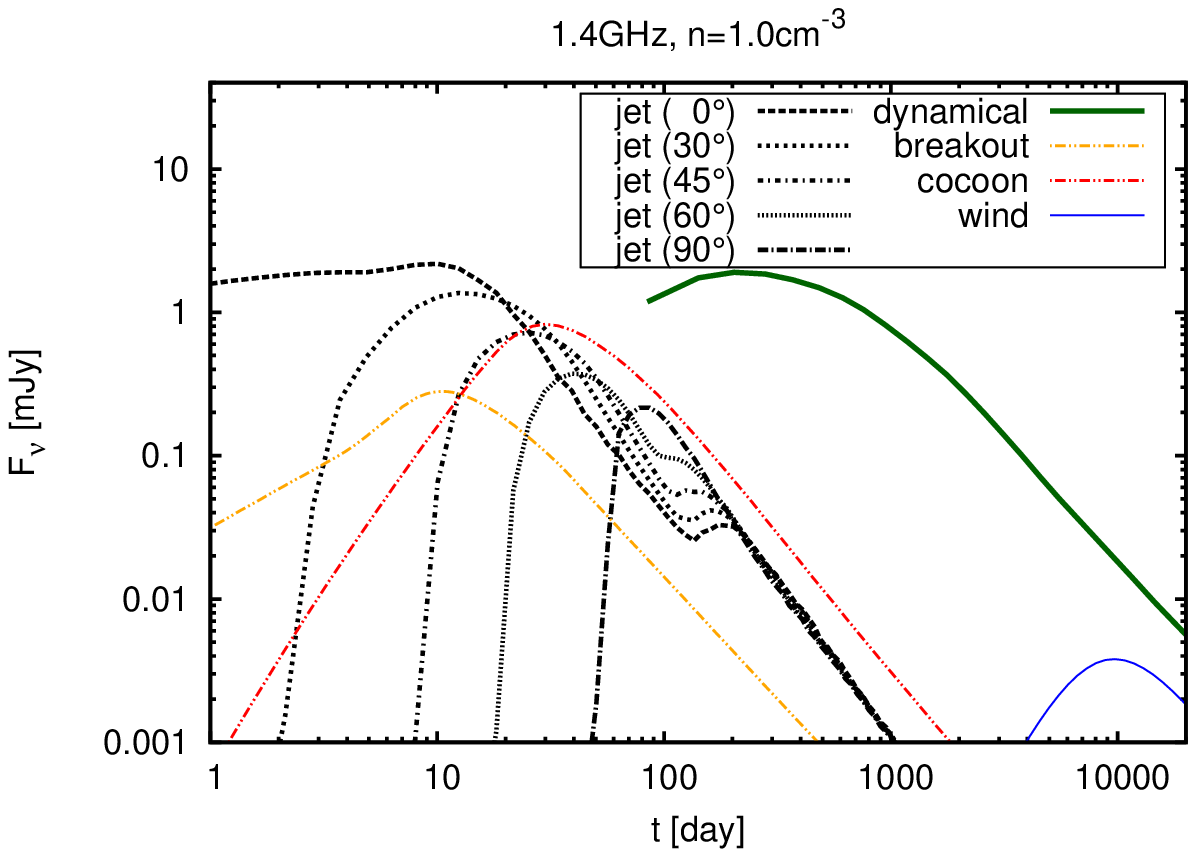}\\
\includegraphics[bb=50 50 410 302,width=85mm]{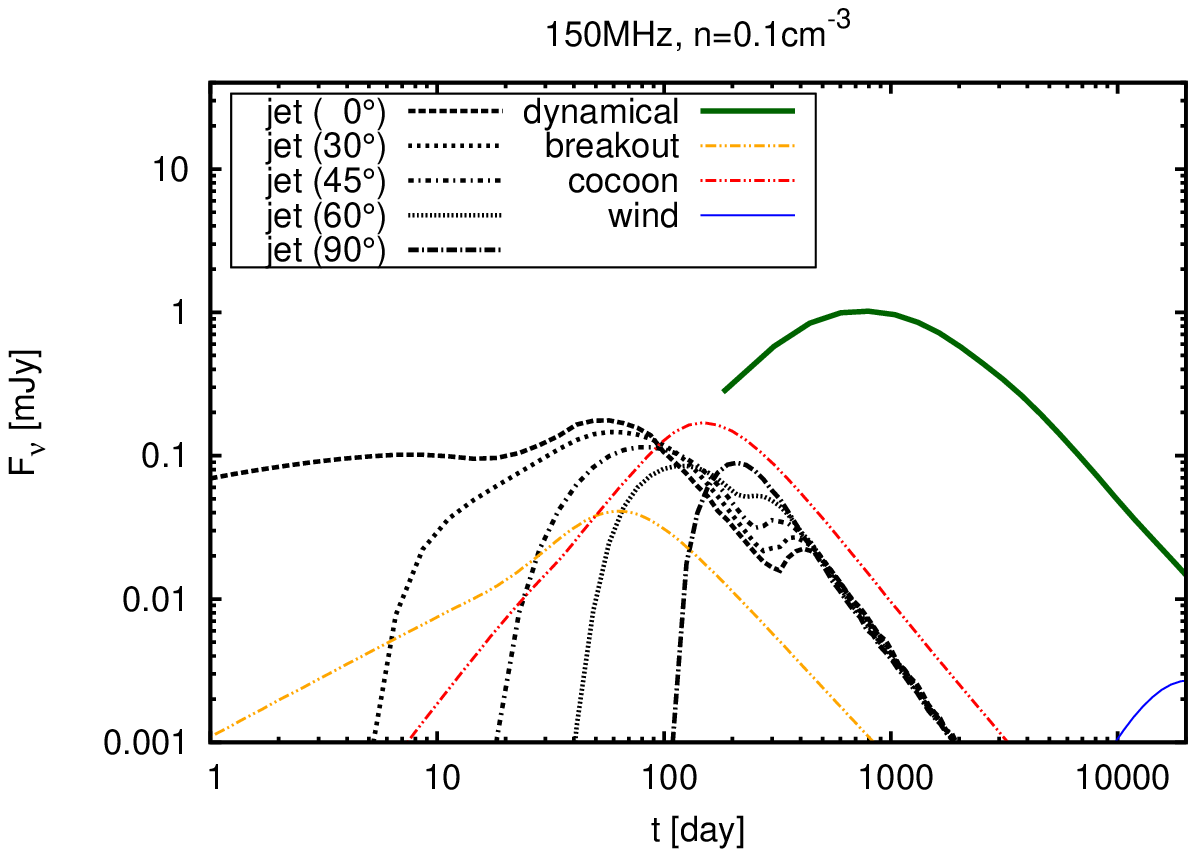}
\includegraphics[bb=50 50 410 302,width=85mm]{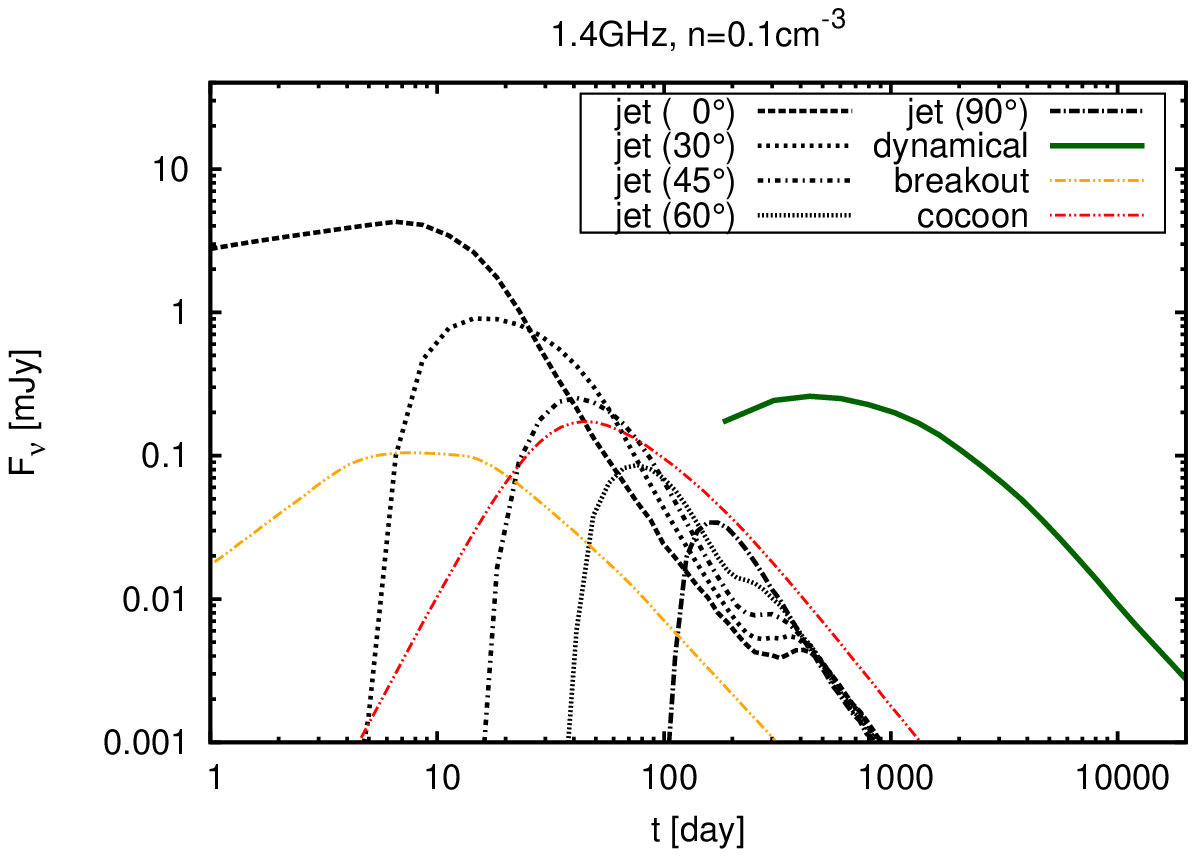}\\
\includegraphics[bb=50 50 410 302,width=85mm]{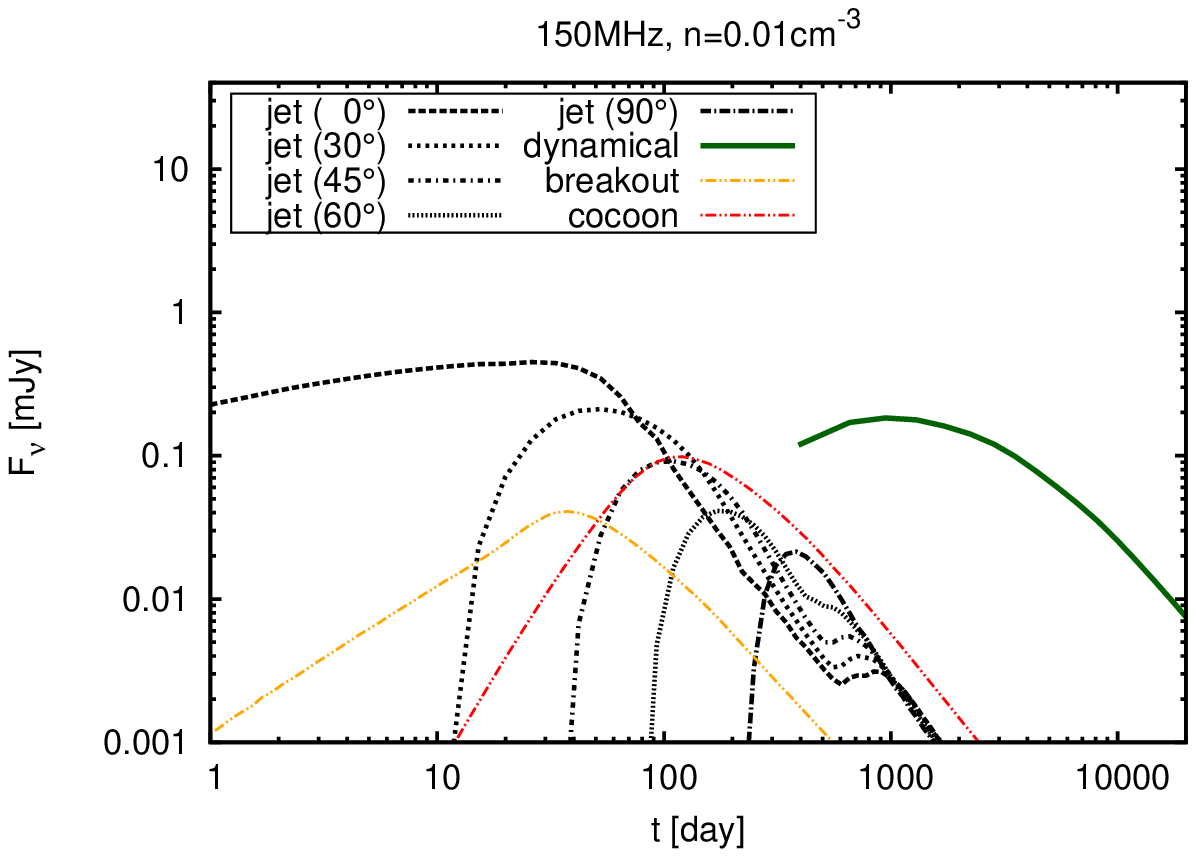}
\includegraphics[bb=50 50 410 302,width=85mm]{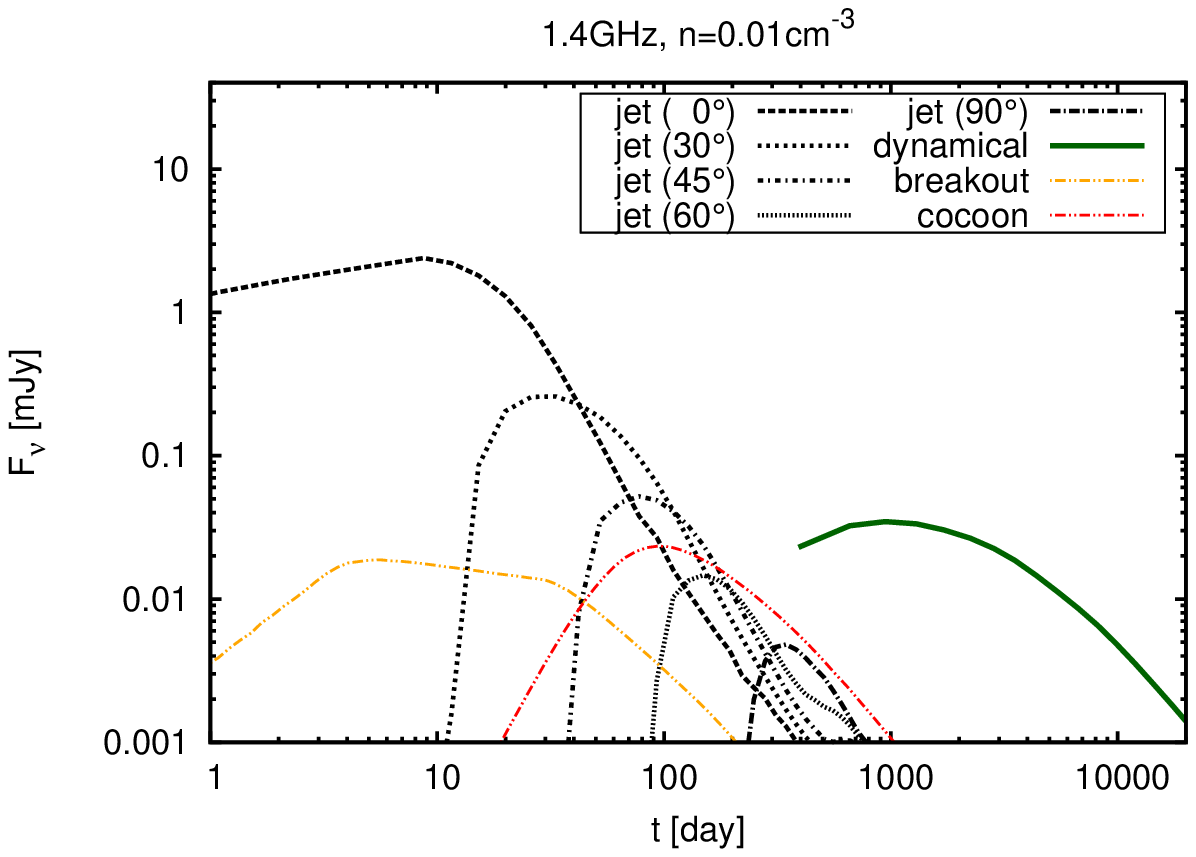}
\caption{
Same as Fig.~\ref{fig2} for the case of a strong GRB.
The kinetic energy of the ultra-relativistic jet and the cocoon is set to be
$10^{49}$~erg. The Lorentz factor of the cocoon is $1.5$~(see Eq.~\ref{Lc}).
}
\label{fig3}
\end{figure*}

Figure~\ref{fig2} depicts the resulting radio light curves
of the different components for our fiducial model~(see Table~\ref{tab1}
for the fiducial parameters). 
We examine three different values of the external density 
$n=0.01$~--~$1~{\rm cm^{-3}}$ and we present the light curves for two observed frequencies  
$150$~MHz (left panels) and  $1.4$~GHz (right panels) 
corresponding to the LOFAR and the EVLA  of radio telescopes.
We set the  luminosity distance of the
source  to be $200$~Mpc, which is roughly the sky averaged
horizon distance of the advanced GW detectors for
ns$^2$ mergers. 

\begin{figure*}
\includegraphics[bb=50 50 410 302,width=85mm]{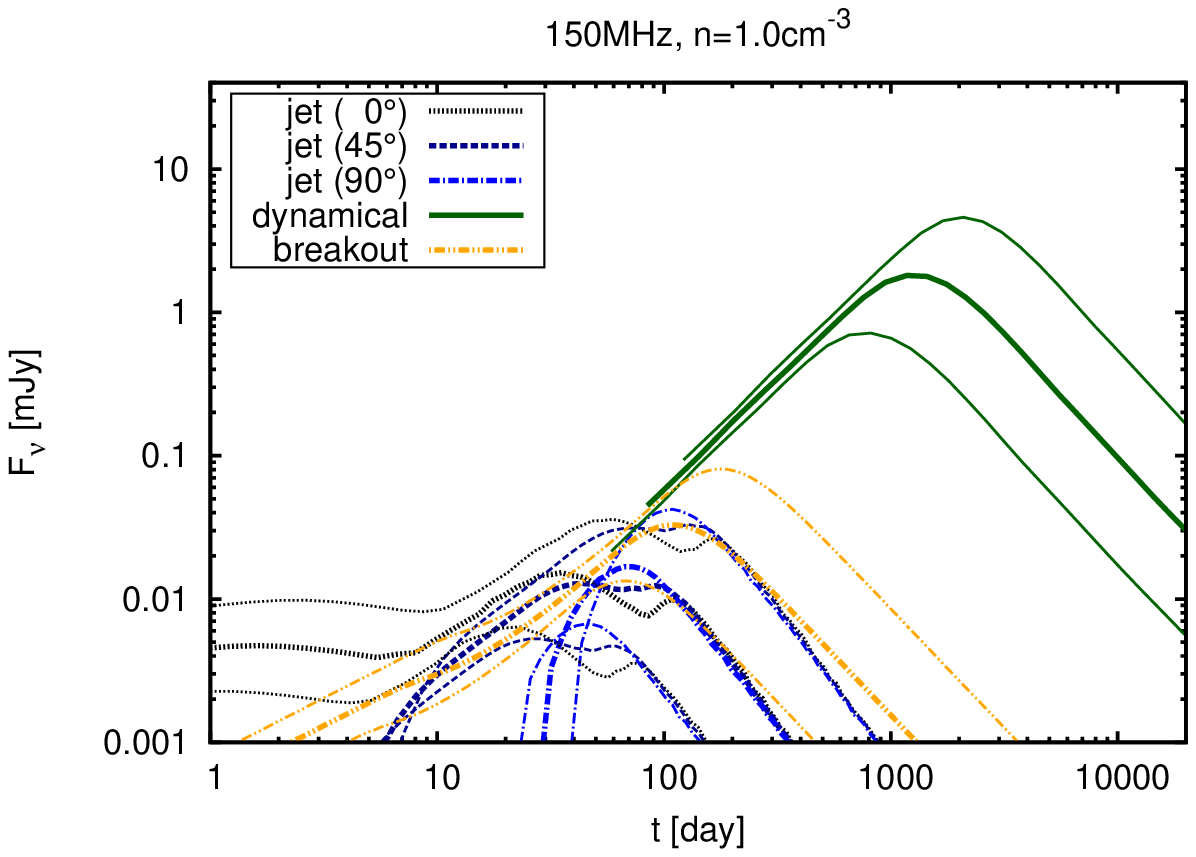}
\includegraphics[bb=50 50 410 302,width=85mm]{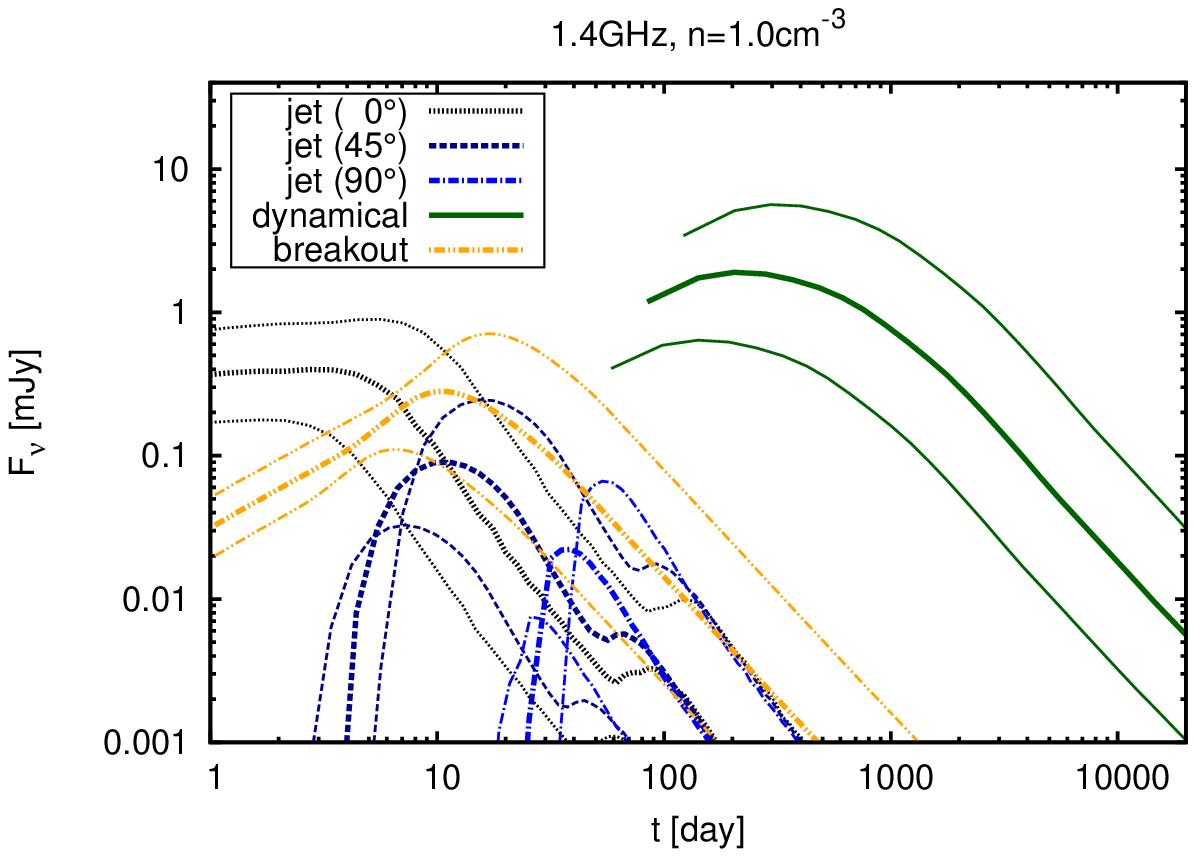}\\
\includegraphics[bb=50 50 410 302,width=85mm]{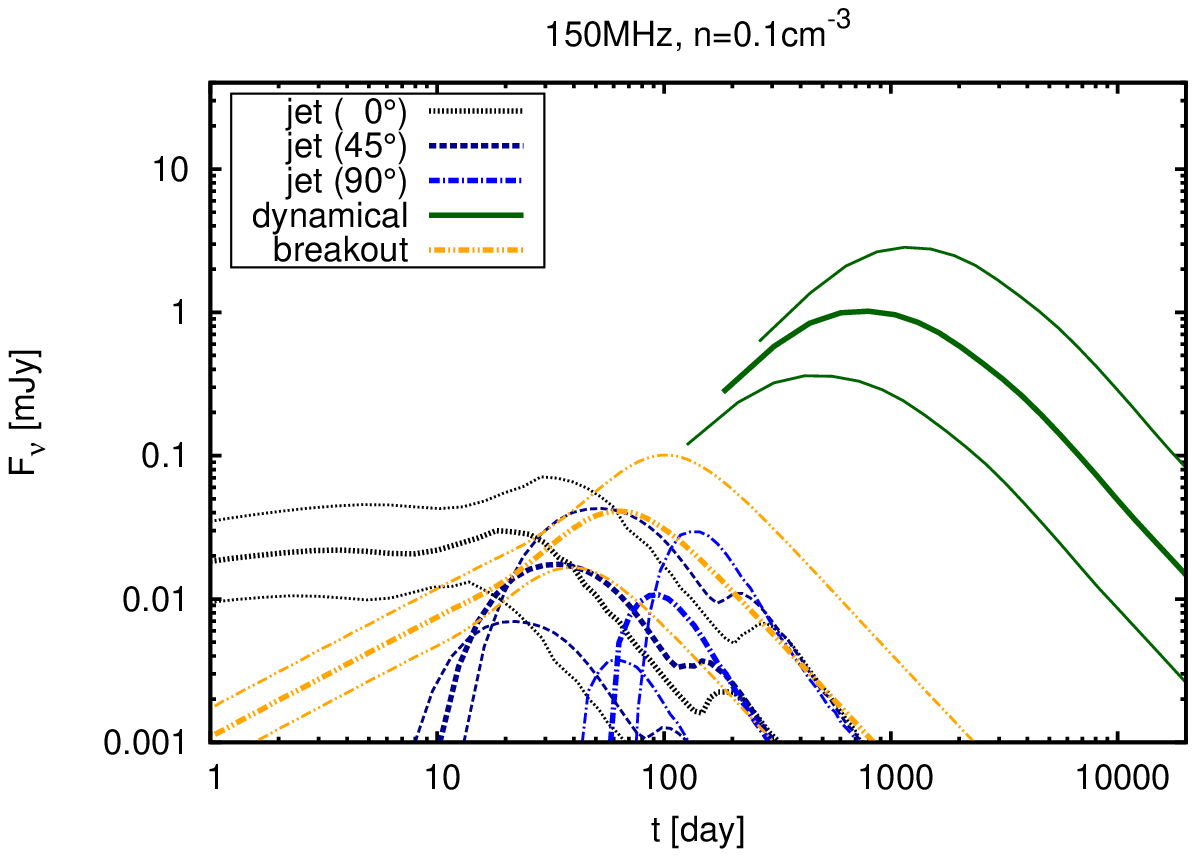}
\includegraphics[bb=50 50 410 302,width=85mm]{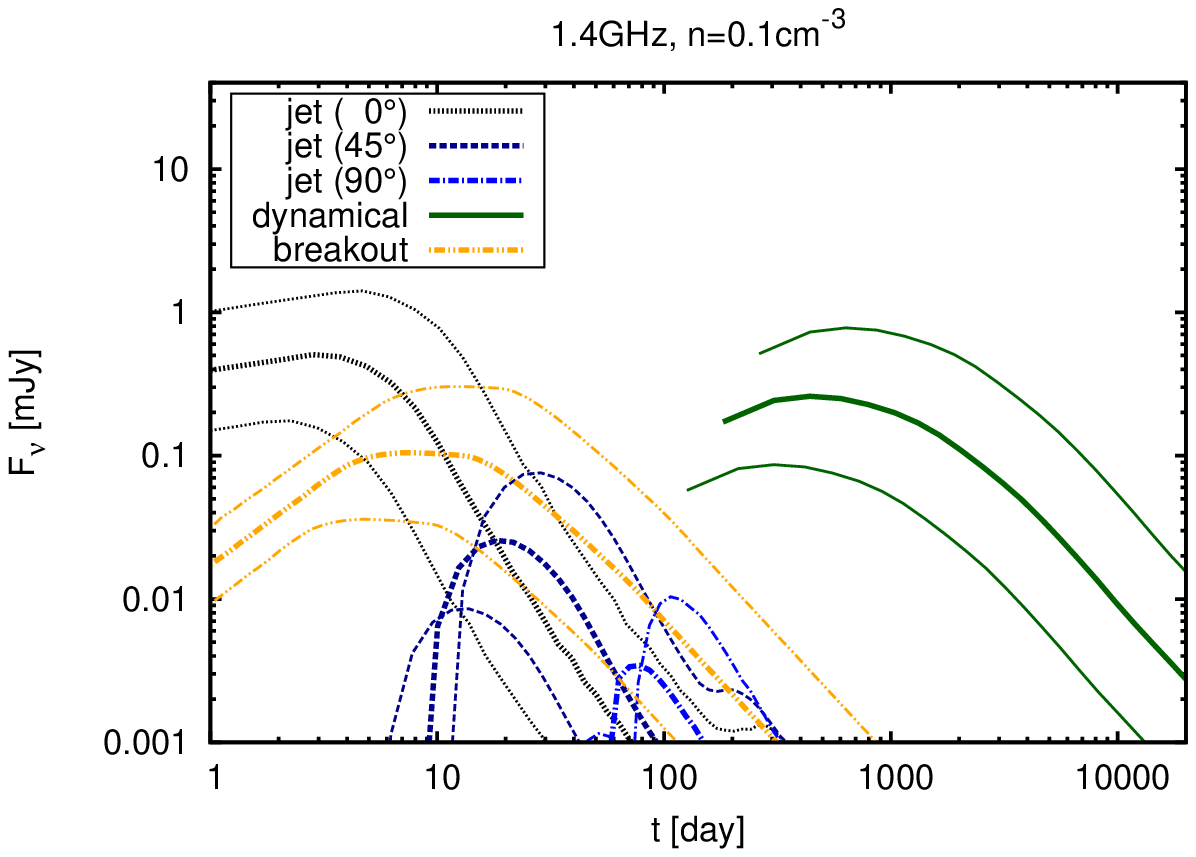}\\
\includegraphics[bb=50 50 410 302,width=85mm]{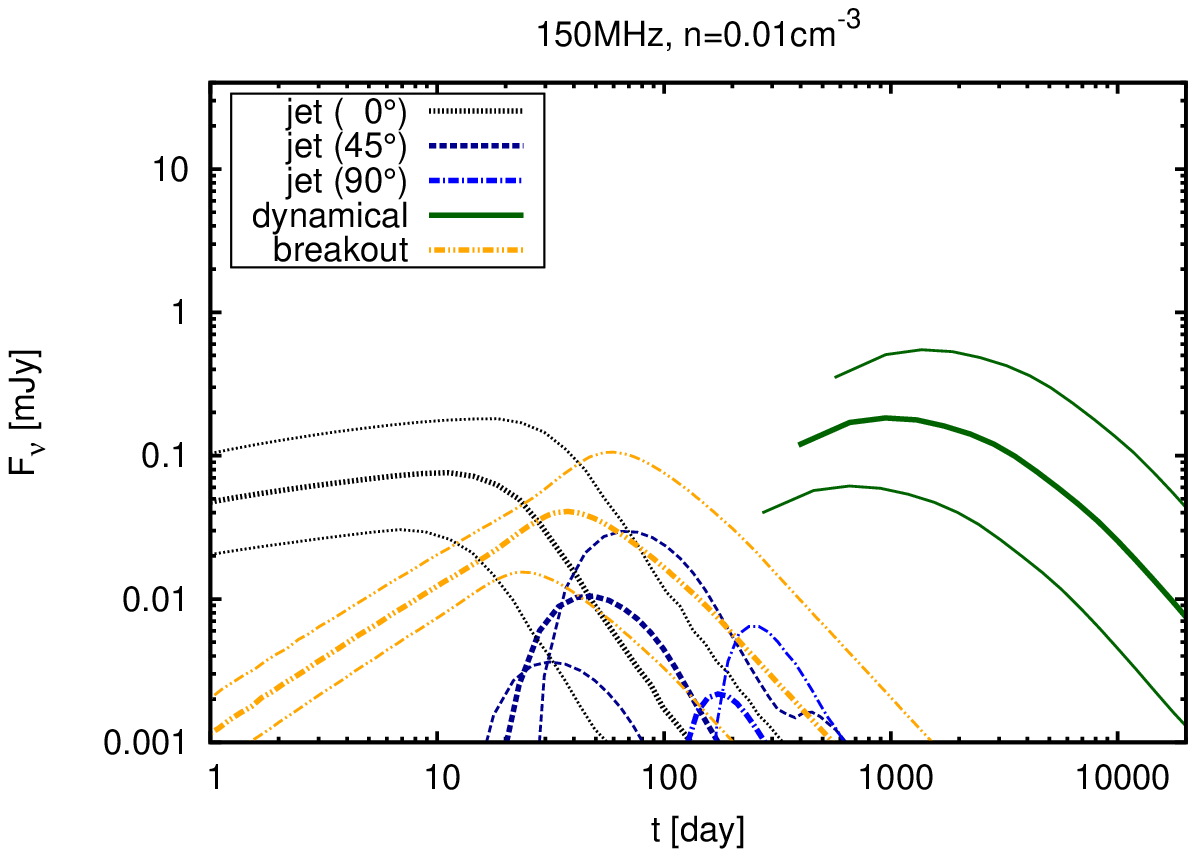}
\includegraphics[bb=50 50 410 302,width=85mm]{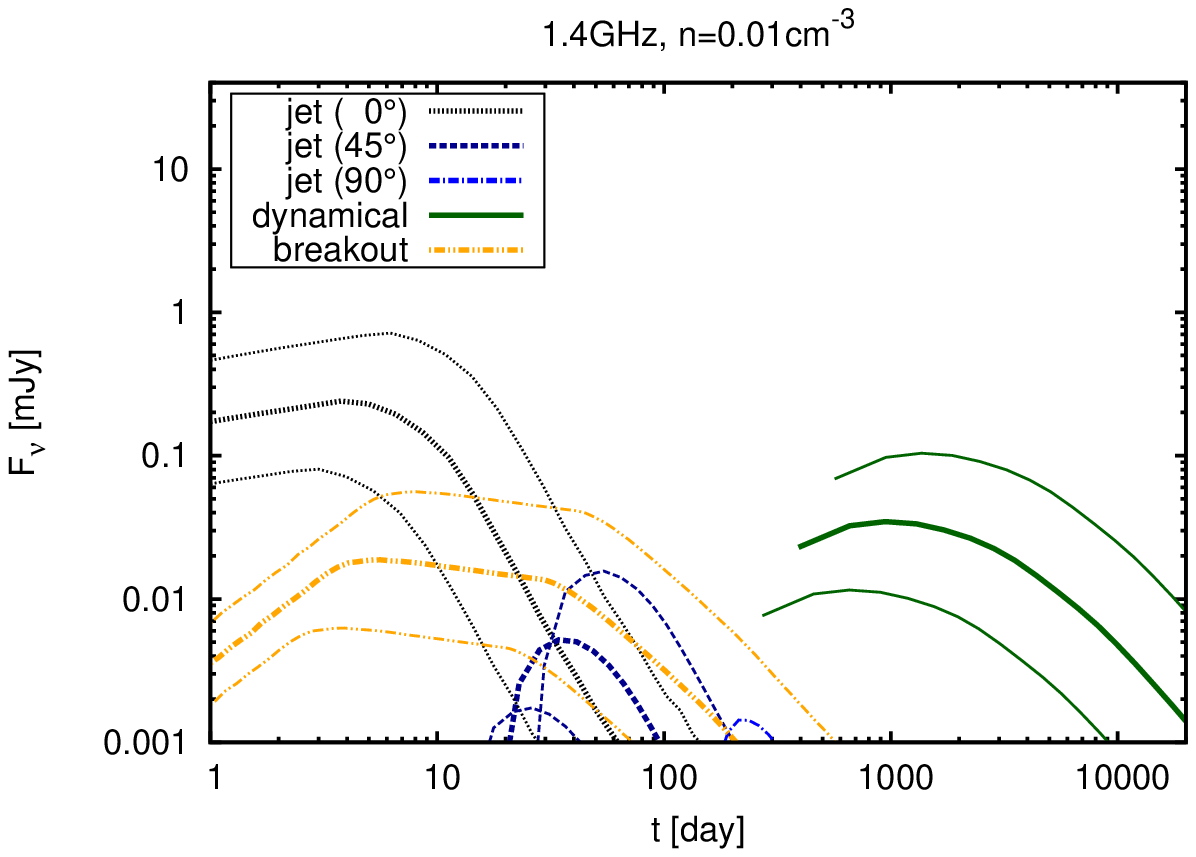}
\caption{
Uncertainties in the radio light curves.
The thick lines denote the light curves of the fiducial case~(see Fig.~\ref{fig2}) and the thin curves denote
the light curves with a kinetic energy larger and smaller by a factor of 3 than
those of the fiducial models.
}
\label{fig4}
\end{figure*}

The ultra-relativistic jet always arrives first. 
This on-axis emission of the jet, the GRB radio afterglow, is the strongest at 1.4 GHz for low external densities $n\lesssim 0.1~{\rm cm^{-3}}$. 
At 150 MHz this GRB afterglow
as well as the other relativistic component, the shock-breakout material, is strongly suppressed by self-absorption and it is much weaker.
For a generic observer,  the off-axis orphan afterglow at viewing angles of 60$^\circ$ and even at 45$^\circ$ 
is always subdominant compared to the shock-breakout material and the dynamical ejecta. 
The mildly relativistic component, that arises from the shock-breakout material, peaks later at around 20--100~days 
depending on the observed frequency and external density. 
Finally the sub-relativistic dynamical ejecta arises at late times~(typically 1000 days).
It is always the brightest at 150 MHz and
it is also brightest at 1.4GHz for higher external densities. 
For our fiducial parameters, that are based on a weak GRB, 
the radio emission from the sub-relativistic cocoon is always negligible.

As mentioned earlier, at early times, 
synchrotron self-absorption strongly suppresses the radio flux at $150$~MHz.
As a result, the peak flux is only
$F_{\nu }\sim 0.01~{\rm mJy}$ for the relativistic components such as
the shock-breakout material and the off-axis GRB jet for all the cases.
As expected from Eq.~(\ref{f2}),  in this case the peak flux depends only on the
kinetic energy among the parameters of ejecta. Indeed, the dynamical
ejecta is the brightest as $F_{\nu}\sim 1~{\rm mJy}$ for $n\gtrsim 0.1~{\rm cm^{-3}}$ and
its peak time is $\sim 1000$~days. 
For very low densities $n \lesssim~0.01 {\rm cm^{-3}}$,
the on-axis GRB afterglow is comparable to the dynamical ejecta flare, 
peaking at about 20~days with $F_{\nu}\sim 0.1 ~{\rm mJy}$.

At $1.4$~GHz, there are the early and late-time radio flares. 
The relativistic components such as the GRB afterglows
and the shock-breakout material contribute the flare 
at early times as expected from Eqs.~(\ref{f1}) and (\ref{t1}).
For low densities $n\lesssim 0.1~{\rm cm^{-3}}$,
the GRB afterglows within a viewing angle $\theta_{\rm obs}\sim 30^{\circ}$
is the brightest at 1.4~GHz as $F_{\nu}\sim 0.1$~--~$0.5$~mJy.
Note that the off-axis GRB afterglows are very
faint for large viewing angle $\theta_{\rm obs}\gtrsim 60^{\circ}$
compared with the shock-breakout material and the dynamical ejecta. At this stage the originally 
beamed jet has already slowed down and its emission is already quasi spherical because of its
low Lorentz factor (this is independent of the question how much did the jet physically expand sideways).
As the jet energy is smaller than those of the other components,
its radiation is weaker. The dynamical component dominates at late times $t\gtrsim 100$~days
and has a relatively flat light curve.

As mentioned earlier, our fiducial GRB was a typical, low luminosity one. 
Figure~\ref{fig3} shows the radio light curves for the case of a strong GRB
with a jet energy of $10^{49}$~erg~(corresponding to an isotropic equivalent energy of $\sim 10^{51}$~erg). 
The energy of the cocoon is also larger as this should be comparable to the 
jet energy. We use  $E_{c}=10^{49}$~erg and a corresponding  Lorentz factor of $\Gamma = 1.5$  is obtained 
from Eq.~(\ref{Lc}). Now, for this GRB, the on-axis GRB afterglow is the brightest at all densities at 1.4~GHz
and for very low densities at 150~MHz. The cocoon is much 
brighter than that of the weak GRB  and its peak flux at 1.4~GHz is comparable to 
that of the dynamical ejecta and to the off-axis orphan afterglow for  $\theta_{\rm obs}=45^{\circ}$.

Unfortunately, there are numerous uncertainties in our estimated light curves. One of the 
strongest sources of the uncertainties arises from lack of precise estimates of the mass and energy 
of the different components. To demonstrate the possible variability of the light curves with these
unknown parameters, we present in  Fig.~\ref{fig4}
the dependence of the radio flares  on the kinetic energy of the different components.
Here we show the light curves of the fiducial model~(thick curves) and those with 
kinetic energies larger and smaller by a factor of 3~(thin curves) than those of
the fiducial model. Above~(below) the self-absorption frequency, the amplitude of the light curves
scales as $\propto E$~($E^{4/5}$) and the timescale of the light curves
behaves as $\propto E^{1/3}$~($E^{5/11}$) as expected (see  Sec.~\ref{sec:mild}).

While the overall peak luminosity and peak flux depend mostly on the global properties 
of the outflow component, the details depend also on 
the spatial distribution and on the velocity distribution. 
For instance, the light curve of an off-axis jet rises steeply because of the
collimation of the jet and relativistic beaming effect. 
The detailed shapes of this rise will depend on the angular structure of the jet. 
The light curves of the spherical
components rise slowly compared to that of off-axis afterglows.
The slope of the light curve depends on the velocity distribution
$E(\geq \Gamma \beta)\propto (\Gamma \beta)^{-\alpha}$. The
rise of light curves will be shallower for lower values of
$\alpha$.

Examination of the numerical simulations reveals that the mildly and sub-relativistic components 
that we have examined do not satisfy indeed the spherical symmetry assumption that we have 
made here.  \cite{margalit2015} 
have estimated the effects of a-sphericity on the emission, focusing
on the dynamical ejecta. They found that, for a given total mass, energy, and external density, 
a-sphericity typically delays the peak emission and reduces the peak flux. 
This can be understood intuitively as follows. If more mass and energy are concentrated 
in one direction the matter propagating in that direction will slow down later.
This longer  deceleration time  results in a longer and weaker radio flare 
compared with the isotropic one. Note that, however, as the outflow is only mildly relativistic,
even from a highly a-spherical ejecta, the emission will be roughly isotropic and viewing angle effects will be small. 
It is worth noting that the effect of a-sphericity is more relevant for 
black hole neutron star mergers, which can result in highly a-spherical mass ejection
(see e.g., \citealt{kyutoku2013PRD,foucart2013PRD}).

\section{The radio signature of the short GRB 130603B}

\label{sec:grb}
\begin{figure}
\includegraphics[bb=50 50 410 302,width=85mm]{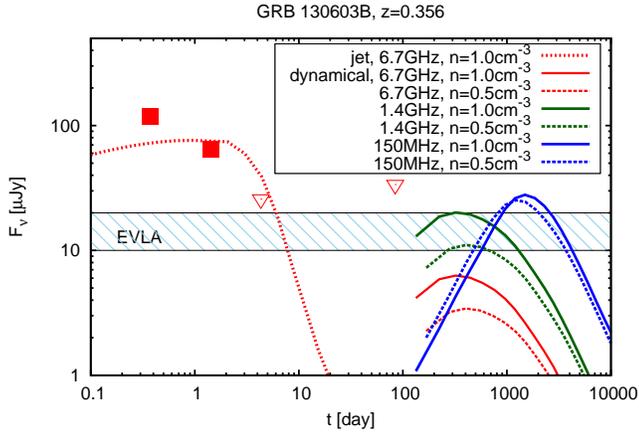}
\caption{
Radio signatures of the short GRB 130603B and light curves at
6.7~GHz~(red curves), 1.4~GHz~(green curves), and 150~MHz~(blue curves).
The red dotted curve denotes the GRB radio afterglow from a jet with
$E=8\times 10^{48}$~erg, $\theta_{j}=4^{\circ}$, $p=2.3$, $\epsilon_{e}=0.2$,
$\epsilon_{B}=8\times 10^{-3}$, and $n=1.0~{\rm cm^{-3}}$. 
The solid and dashed curves denote the expected radio
light curves from a dynamical ejecta with an external density $n=1.0~{\rm cm^{-3}}$
and $0.5~{\rm cm^{-3}}$, respectively. For the dynamical ejecta, the kinetic energy is assumed
to be $8\times 10^{50}$~erg and the other microphysics parameters are the same as
those in Sec.~\ref{sec:radio}. The filled squares and the open triangles show the
observed data points and the upper limits at 6.7~GHz obtained with the VLA~\citep{fong2014ApJ}.
The blue shaded region shows the expected sensitivity of the EVLA at 1.4~GHz.}
\label{fig5}
\end{figure}

The short GRB 130603B had an associated  macronova candidate
\citep{berger2013ApJ,tanvir2013Nature}.
While the macronova identification is based only on
one observed data point
in the $H$-band at about $7$~days~(in source frame) after the burst, this is 
the first, even though rather weak, evidence of the significant mass ejection from a ns$^{2}$.
Using this data point, one can  estimate,
from the observed luminosity of the macronova, the minimal ejecta mass  as
$M_{\rm ej}\approx 0.02(\epsilon_{\rm th}/0.5)~M_{\odot}$~\citep{hotokezaka2013ApJ,piran2014},
where $\epsilon_{\rm th}$ is the conversion efficiency
from the total energy generated by radioactive decay into
the thermal energy of the ejecta. The velocity can be also estimated
as $v\gtrsim 0.1c$ from the condition that radiation can diffuse out
from the ejecta with a mass $M_{\rm ej}\gtrsim 0.02M_{\sun}$ 
and an opacity of $10~{\rm cm^{2}/g}$
\citep{kasen2013ApJ,tanaka2013ApJ} at about $7$~days after the burst.
Assuming that the ejecta mass and average velocity is 
$M_{\rm ej}\sim 0.02M_{\odot}$ and $v\sim 0.2c$,
the estimated kinetic energy is about $10^{51}$~erg.

Figure~\ref{fig5} shows 
the expected radio flares from the dynamical ejecta at 150MHz, 1.4~GHz, and 6.7~GHz
as well as the observed early radio afterglow of GRB 130603B at 6.7~GHz by the VLA~\citep{fong2014ApJ}. 
Here we also show a GRB afterglow light curve which is obtained with parameters
$E=8\times 10^{48}$~erg, $\theta_{j}=4^{\circ}$, $p=2.3$, $\epsilon_{e}=0.2$,
$\epsilon_{B}=8\times 10^{-3}$, and $n=1.0~{\rm cm^{-3}}$. The light curve is consistent with 
the observed data points and upper limits\footnote{The light curve that we obtain 
is also consistent with the observed data in other frequencies  
except for the late time excess in the $H$-band~(the macronova candidate) and in the X-ray band.} 
within a factor of 2.
It is worth emphasizing that   
the parameters can change by orders of magnitude and still fit the data. For instance, the external density
lies in the range $n\approx 5\times 10^{-3}$~--~$30~{\rm cm^{-3}}$~\citep{fong2014ApJ}. 
For modeling the radio flare from the dynamical ejecta, we assume an external density to be $1.0$ and $0.5~{\rm cm^{-3}}$
and we use an ejecta mass $M=0.02M_{\odot}$ and a kinetic energy $E=8\times 10^{50}$~erg.
The predicted dynamical ejecta light curves at 6.7~GHz are well below the upper limits 
$F_{\nu}\lesssim 30~\mu{\rm Jy}$ at 
$\sim 80$ days. 
However, later observations may detect a signal. 
Specifically the peak flux at 1.4~GHz can be as high as $F_{\nu}\approx 20~\mu{\rm Jy}$ 
(depending on the external density). The expected sensitivity of the EVLA
at 1.4~GHz is also shown in the figure.
For the higher range of external densities $n\gtrsim 0.5~{\rm cm^{-3}}$, 
the radio flare might be detectable with the EVLA.
The signal at 150~MHz can be $F_{\nu}\approx 30~\mu{\rm Jy}$, which might be also detectable
with the LOFAR. 
A positive detection of a varying radio signal will confirm the identification of
this event as a ns$^2$ merger and will establish the observed infrared bump as a macronova.

\section{Conclusion and discussion}
\label{sec:conc}

A ns$^2$ merger ejects a significant amount of mass in several  different components:
a dynamical ejecta,
a shock-breakout material,
a wind from a black hole/neutron star surrounded by
an accretion disk, and a relativistic jet, producing a GRB. 
As a result of the interaction of the relativistic jet with
the earlier and slower ejecta along the rotational axis,
a cocoon is expected to be formed and expands nearly spherically.
Among the different components of the ejecta,
the dynamical ejecta, which is also the most robustly found in numerical simulations, has the largest amount of kinetic energy up to $E\sim 10^{51}$~erg.
This is comparable to the ``isotropic equivalent'' energy of the highly beamed GRB jet.
Somewhat surprisingly the beamed GRB jet is among the least energetic.
Overall we have three important distinct components,
the dynamical ejecta, which is mostly sub-relativistic, 
the mildly relativistic shock-breakout material and cocoon
\footnote{Note that the velocity of the cocoon
depends on the energy deposited into the cocoon. The cocoon will be
relativistic when the deposited energy is larger than about $10^{49}$~erg.},
and the ultra-relativistic beamed jet. 

We have calculated the expected radio signals produced
via synchrotron emission from electrons accelerated
in the shocks formed between the different components of the ejecta
and the ISM. This would be a low frequency (as opposed to the X-ray or optical GRB afterglow 
or the optical--infrared macronova) electromagnetic counterpart
of the GW event. This process is similar 
to GRB afterglows and radio emission
of some early supernova remnants. 
In contrast with the high frequency counterparts, this emission lasts much longer and 
may even peak a few years after the merger. 
We focused on the expected
radio flux at two frequencies of 150~MHz and 1.4~GHz.
We found that there are three types of the radio flares: (i) The ultra-relativistic
jet produces the earliest bright radio flare with a timescale of $\sim 10$~days for an
observer along the jet axis or close to it. (ii) A radio flare with a timescale
of a few dozen days  is produced by the mildly relativistic components such as 
the shock-breakout material, the off-axis jet, and the cocoon. The latter flare from  
the cocoon is significant only when the cocoon is energetic enough so that it has a relativistic velocity. 
(iii) Finally the sub-relativistic dynamical ejecta produces a  radio flare with a timescale of a few years.
At 150~MHz, however, the radio flare is strongly suppressed by
synchrotron self absorption for $n\gtrsim 0.01~{\rm cm^{-3}}$ until a few months. 
Hence the earlier signals are much weaker at low frequencies. 
Depending on the external density and the distance to the source,
these radio flares could be detectable as GW counterparts. Such a detection 
can reveal the nature of ns$^{2}$ mergers.  

Although the modeling of radio emission from mergers
contains uncertain parameters such as the kinetic energy of the ejecta
and the external density, 
it is worth to estimate these values from the nature of the short GRB 130603B.
The detection of a macronova candidate
associated with this event allows us
to estimate the ejecta mass $M_{\rm ej}\approx 0.02(\epsilon_{\rm th}/0.5)~M_{\odot}$.
Assuming the velocity of the ejecta is $0.2c$,
the estimated kinetic energy is about $10^{51}$~erg.
The detection of the afterglows implies that the external density is in the range of 
$n\approx 5\times 10^{-3}$~--~$30~{\rm cm^{-3}}$~\citep{fong2014ApJ}. 
The radio afterglow of GRB 130603B, that arose from the relativistic jet,  decayed quickly and it was below $30~\mu {\rm Jy}$ 
at $\sim 4$ days with a similar upper limit at 80~days. 
However it is still possible to observe the late-long lasting radio signal arising from the dynamical ejecta.  
This signal could  be as high as $20~\mu{\rm Jy}$  at 1.4~GHz depending 
mostly on the external density.
For the higher range of external densities $n\gtrsim 0.5~{\rm cm^{-3}}$, this would be detectable at
1.4~GHz with the EVLA. At 150~MHz, the expected flux is about $30~\mu{\rm Jy}$, which depends weakly on
the external density and peaks at late times. This flux might be detectable with the LOFAR.
While a detection is uncertain, a positive radio signal will confirm 
the identification of this event as a ns$^2$ merger and will establish the observed infrared
bump as a macronova.

\vspace{-0.5cm}
\section*{Acknowledgments}
We thank J. Granot, E. Nakar, L. Nava, S. Nissanke, 
R. Sari, and R. Shen for fruitful discussions.
This research was supported by an ERC advanced grant (GRBs) and by the  I-CORE 
Program of the Planning and Budgeting Committee and The Israel Science
Foundation (grant No 1829/12). 

\bibliographystyle{mn2e}

\end{document}